\documentclass[paper=A4,11pt,DIV=12,headings=big]{scrartcl}
\pdfoutput=1
\usepackage{graphicx}
\usepackage{tabularx}
\usepackage{array}
\usepackage{amsmath}
\usepackage{amssymb}
\usepackage{xcolor}
\usepackage{longtable}
\usepackage{verbatim}
\usepackage{cite}
\usepackage{bbm}
\usepackage{multirow}
\usepackage[normalem]{ulem}
\definecolor{darkblue}{rgb}{0,0,0.5}
\usepackage[colorlinks,linkcolor=darkblue,citecolor=darkblue,urlcolor=darkblue,%
pdftitle={New physics in B->K*mumu},%
pdfauthor={Wolfgang Altmannshofer, David M. Straub}]{hyperref}

\allowdisplaybreaks

\addtokomafont{disposition}{\rmfamily\boldmath}

\newcommand{\mailref}[1]{\href{mailto:#1}{#1}}

\newcommand{\beq}{\begin{equation}}
\newcommand{\eeq}{\end{equation}}
\newcommand{\ba}{\begin{array}}
\newcommand{\ea}{\end{array}} 
\newcommand{\beqa}{\begin{eqnarray}}
\newcommand{\eeqa}{\end{eqnarray}}

\def \eff{\rm eff}

\begin{document}

\begin{flushright}
\begin{tabular}{l}
\small
FERMILAB-PUB-13-310-T\\
MITP/13-047
\end{tabular}
\end{flushright}
\vskip0.5cm

\begin{center}
{\Huge\color{black!70}
\bf\boldmath\textsl{New Physics in} $B \to K^*\!\mu\mu$?}\\[0.8 cm]
{\large
Wolfgang~Altmannshofer$^{a}$ and David M. Straub$^{b}$} \\[0.4 cm]
\small
$^a$ {\em Fermi National Accelerator Laboratory, P.O. Box 500, Batavia, IL 60510, USA}\\[0.1cm]
$^b${\em PRISMA Cluster of Excellence \& Mainz Institute for Theoretical Physics, \\
Johannes Gutenberg University, 55099 Mainz, Germany} \\[0.4cm]
E-mail: \mailref{waltmann@fnal.gov}, \mailref{david.straub@uni-mainz.de}%
\end{center}

\medskip
\begin{abstract}\noindent
Recent experimental results on angular observables in the rare decay $B\to K^*\mu^+\mu^-$ show significant deviations from Standard Model predictions. We investigate the possibility that these deviations are due to new physics. Combining all relevant data on $b \to s$ rare decays, we show that a consistent explanation of most anomalies can be obtained by new physics contributing simultaneously to the semi-leptonic vector operator $O_9$ and its chirality-flipped counterpart $O_9'$.
A partial explanation is possible with new physics in $O_9$ or in dipole operators only.
We study in detail the implications for models of new physics, in particular the minimal supersymmetric standard model, models with partial compositeness and generic models with flavour-changing $Z^\prime$ bosons. In all considered models, contributions to $B\to K^*\mu^+\mu^-$ of the preferred size imply a spectrum close to the TeV scale. 
We stress that measurements of CP asymmetries in $B\to K^*\mu^+\mu^-$ could provide valuable information to narrow down possible new physics explanations.
\end{abstract}

\setcounter{tocdepth}{2}
\tableofcontents

\section{Introduction}

Rare $B$ decays mediated by the flavour-changing neutral $b\to s$ transition are sensitive probes of physics beyond the Standard Model (SM), being loop and CKM-suppressed in the SM. In the LHC era, particular interest lies on exclusive semi-leptonic and leptonic decays, since they can be measured more readily in a hadron collider environment. The decay $B\to K^*(\to K\pi)\mu^+\mu^-$ plays a special role since the angular distribution of its four-body final state gives access to numerous observables sensitive to new physics (NP).
Since, for a neutral $B$ decay, the flavour can be tagged by measuring the charges of the final-state mesons, it is also straightforward to measure CP asymmetries in the angular distribution, some of which are not suppressed by small strong phases \cite{Bobeth:2008ij}.
All of this makes this decay mode a unique laboratory to test the size, chirality structure and CP phase of the FCNC transition in the SM and beyond, as has been discussed extensively in the literature (see e.g.\ \cite{Bobeth:2010wg,Alok:2010zd,Alok:2011gv,DescotesGenon:2011yn,Bobeth:2011gi,Becirevic:2011bp,Altmannshofer:2011gn,Bobeth:2011nj,Matias:2012xw,Beaujean:2012uj,Altmannshofer:2012az,Becirevic:2012dx,DescotesGenon:2012zf,Bobeth:2012vn,Descotes-Genon:2013vna} for recent studies).

After initial measurements of branching ratio and angular observables at $B$ factories~\cite{Wei:2009zv,Lees:2012tva} and the Tevatron~\cite{Aaltonen:2011qs,Aaltonen:2011ja,CDFupdate}, recent measurements by LHCb, ATLAS and CMS \cite{Aaij:2013iag,Aaij:2013qta,ATLAS:2013ola,CMS:cwa}
have added a wealth of new data. In particular, the recent measurement presented by the LHCb collaboration \cite{Aaij:2013qta}
shows several significant deviations from the SM expectations. While it is conceivable that these anomalies are due to statistical fluctuations or underestimated theory uncertainties -- for recent reappraisals of various sources of uncertainty, see \cite{Khodjamirian:2010vf,Beylich:2011aq,Becirevic:2012dp,Matias:2012qz,Jager:2012uw} -- here we investigate the possibility that they are due to new physics.
Finding a consistent explanation in terms of new physics is highly non-trivial since all the observables in $B\to K^*\mu^+\mu^-$ -- many of which are in agreement with SM expectations -- as well as other decays like $B_s\to\mu^+\mu^-$, $B\to K\mu^+\mu^-$ or $B\to X_s\gamma$, depend on the same short-distance Wilson coefficients, such that a global analysis of model-independent constraints is required. We perform such analysis, building on our previous work
\cite{Altmannshofer:2011gn,Altmannshofer:2012az}, with refinements detailed below.
Very recently, ref.~\cite{Descotes-Genon:2013wba} appeared, that also performs a model independent analysis of $B \to K^* \mu^+\mu^-$ anomalies. We will compare our findings in the conclusions.

Our paper is organized as follows.
In sections~\ref{sec:Heff} and \ref{sec:obs} below, we define the relevant effective Hamiltonian and briefly discuss the observables in $B\to K^*\mu^+\mu^-$.
In section~\ref{sec:fit}, we perform a model independent analysis of NP effects in $b\to s$ transitions, identifying the Wilson coefficients whose modification can lead to a consistent explanation of the experimental observations.
In section~\ref{sec:NPmodels}, we discuss three concrete NP models: a model with a heavy neutral gauge boson ($Z'$), the MSSM, and models with partial compositeness.
Section~\ref{sec:concl} contains our conclusions.

\subsection{Effective Hamiltonian}\label{sec:Heff}

The effective Hamiltonian for $b\to s$ transitions can be written as
\begin{equation}
\label{eq:Heff}
{\cal H}_{\eff} = - \frac{4\,G_F}{\sqrt{2}} V_{tb}V_{ts}^* \frac{e^2}{16\pi^2}
\sum_i
(C_i O_i + C'_i O'_i) + \text{h.c.}
\end{equation}
and we consider NP effects in the following set of dimension-6 operators,
\begin{align}
O_7^{(\prime)} &= \frac{m_b}{e}
(\bar{s} \sigma_{\mu \nu} P_{R(L)} b) F^{\mu \nu},
&
O_9^{(\prime)} &= 
(\bar{s} \gamma_{\mu} P_{L(R)} b)(\bar{\ell} \gamma^\mu \ell)\,,
&
O_{10}^{(\prime)} &=
(\bar{s} \gamma_{\mu} P_{L(R)} b)( \bar{\ell} \gamma^\mu \gamma_5 \ell)\,.
\label{eq:ops}
\end{align}
We ignore scalar or pseudoscalar operators, since they are numerically irrelevant for the $B\to K^*\mu^+\mu^-$ decay. In models modifying $O_7^{(\prime)}$, typically also the chromomagnetic penguin operator $O_8^{(\prime)}$ is modified. However, since it enters the decays considered below only via operator mixing with  $O_7^{(\prime)}$, its discussion is redundant.
As in our previous studies, in our numerical analysis we consider NP effects to $C_7^{(\prime)}$ at a matching scale of $160\,\text{GeV}$.

\subsection{Observables}\label{sec:obs}

In general, the angular distribution of $B\to K^*\mu^+\mu^-$ contains 24 observables \cite{Altmannshofer:2008dz},
expressed as angular coefficients of a three-fold differential decay distribution, that are functions of the dilepton invariant mass squared $q^2$. However, setting the lepton mass to zero (which is justified at the current level of experimental precision) and neglecting effects of the scalar and pseudoscalar operators (which is well-motivated for $B\to K^*\mu^+\mu^-$ due to the absence of large enhancements in $B_s\to\mu^+\mu^-$) one is left with 18 independent observables.
A convenient basis for these 18 observables, reducing theoretical uncertainties and separating CP-violating from CP-conserving effects was suggested in \cite{Altmannshofer:2008dz}. Instead of the angular coefficients of the $B$ and $\bar B$ decay, one considers their sum or difference, normalized to the differential decay rate,
\begin{equation}
 S_i = \left( I_i + \bar I_i \right) \bigg/ \frac{d(\Gamma+\bar\Gamma)}{dq^2} \,,
\qquad
 A_i = \left( I_i - \bar I_i \right) \bigg/ \frac{d(\Gamma+\bar\Gamma)}{dq^2}\,.
\label{eq:As}
\end{equation}
Binned observables, defined as ratios of $q^2$ integrals of numerator and denominator, are denoted as $\langle S_i \rangle_{[a,b]}$ and $\langle A_i \rangle_{[a,b]}$.

\begin{table}
\centering
\renewcommand{\arraystretch}{1.4}
\begin{tabular}{ccccccc}
\hline\hline
\textbf{Here}  & \cite{Altmannshofer:2008dz} & \cite{Bobeth:2008ij} & \cite{Descotes-Genon:2013vna} & LHCb \cite{Aaij:2013iag,Aaij:2013qta}
 & sens. at low $q^2$ & sens. at high $q^2$ \\
\hline
\boldmath $F_L$ & $-S_2^c$ & & $F_L$ & $F_L$ &
$C_{7,9}, C_{9,10}'$ & $C_{9,10}'$\\
\boldmath $A_\text{\textbf{FB}}$ & $\frac{3}{4}S_6^s$ & $A_\text{FB}$ & $-A_\text{FB}$ & $-A_\text{FB}$ &
$C_7,C_9$ & $C_{9,10}, C_{9,10}'$ \\
\boldmath $S_3$ & $S_3$ &&$\frac{1}{2}F_T\,P_1$& $S_3$& 
$C_{7,10}'$ & $C_{9,10}'$\\
\boldmath $S_4$ & $S_4$ &  & $\frac{1}{2}F_{LT}\,P_4'$ & $-S_4$ & $C_{7,10},C_{7,10}'$ & $C_{9,10}'$\\
\boldmath $S_5$ & $S_5$ &  & $F_{LT}\,P_5'$&  $S_5$& $C_{7,9},C_{7,9,10}'$& $C_9, C_{9,10}'$ \\
\boldmath $A_7$ & $A_7$ & $-\frac{2}{3}A_7^D$ & $-F_{LT}\,{P_6'}^\text{CP}$& &
$C_{7,10},C_{7,10}'$ & -- \\
\boldmath $A_8$ & $A_8$ & $-\frac{2}{3}A_8^D$ &  $-\frac{1}{2}F_{LT}\,{P_8'}^\text{CP}$ &&
$C_{7,9},C_{7,9,10}'$ & $C_{9,10}'$\\
\boldmath $A_9$ & $A_9$ & $\frac{2}{3}A_9$ &  & $A_9$ & 
 $C_{7,10}'$ & $C_{9,10}'$\\
\hline
&&& $P_4'$ & $-2 P_4'$ \\
&&& $P_5'$ & $P_5'$ \\
\hline\hline
\end{tabular}
\caption{Dictionary between different notations for CP averaged angular coefficients and CP asymmetries in $B \to K^* \mu^+\mu^-$, where $F_{LT}\equiv\sqrt{F_LF_T}$ and $F_T=1-F_L$. The last two columns show the Wilson coefficients the observables are most sensitive to both at low $q^2$ and high $q^2$.}
\label{tab:dict}
\end{table}

Only some of these angular observables are sensitive to new physics effects in the operators (\ref{eq:ops}) though. In addition to the differential decay rate -- which is subject to sizable theory uncertainties -- there are mainly five CP-averaged angular observables and three CP-asymmetries that can receive significant NP contributions.
We list them
in table~\ref{tab:dict} and compare them to other conventions used in the literature.
We also compare our conventions to the set of ``optimized'' observables suggested in \cite{Descotes-Genon:2013vna}. These observables correspond to the $S_i$ and $A_i$ divided by a function of the $K^*$ longitudinal polarization 
fraction $F_L(q^2)$ and are constructed to reduce the dependence on hadronic form factors.
In the last two columns, we list the Wilson coefficients that can lead to visible
NP effects in the observables in question. Here one has to distinguish between the low $q^2$ region, $q^2\lesssim8\,\text{GeV}^2$, and the high-$q^2$ region, $q^2\gtrsim14\,\text{GeV}^2$. The intermediate region is unreliable due to the presence of charmonium resonances.

\section{Anatomy of new physics effects and fit to the data}\label{sec:fit}

The methodology of our global analysis of constraints on Wilson coefficients is based on our two previous studies \cite{Altmannshofer:2011gn,Altmannshofer:2012az} and described there in detail. Here we only list the changes in the experimental input data and theoretical calculations with respect to \cite{Altmannshofer:2012az}.
On the theory side, we use the recent lattice calculation of $B\to K$ form factors at high $q^2$ by the HPQCD collaboration \cite{Bouchard:2013eph,Bouchard:2013mia}, which strongly reduces theoretical uncertainties. On the experimental side, we now additionally include
\begin{itemize}
\item updates of the angular analysis of $B^0\to K^{*0}\mu^+\mu^-$ by LHCb \cite{Aaij:2013iag,Aaij:2013qta}, 
ATLAS \cite{ATLAS:2013ola} and CMS \cite{CMS:cwa},
\item an update of $\text{BR}(B^+\to K^+\mu^+\mu^-)$\footnote{%
We do not use the LHCb measurement of $\text{BR}(B^0\to K^0\mu^+\mu^-)$ \cite{Aaij:2012cq}, which has larger error bars. Note that this measurement is on the low side compared to the charged decay, which is hard to accommodate even in the presence of new physics \cite{Lyon:2013gba}. We do not average the $B^+\to K^+\mu^+\mu^-$ data with $B$ factory or CDF data either, since they have a numerically negligible impact.
}
by LHCb \cite{Aaij:2012vr},
\item the recent measurements of $\text{BR}(B_s\to\mu^+\mu^-)$ by CMS \cite{Chatrchyan:2013bka} and LHCb \cite{Aaij:2013aka}.
\end{itemize}
Recently, the LHCb collaboration also found an unexpectedly large contribution of a charmonium resonance in the high dilepton invariant mass region of the decay $B^+\to K^+\mu^+\mu^-$ that makes up $\sim 20\%$ of the $B^+\to K^+\mu^+\mu^-$ signal yield~\cite{Aaij:2013pta}. To be conservative, we treat this result by adding an additional relative theoretical uncertainty of 20\% to all high-$q^2$ observables, both in $B\to K\mu^+\mu^-$ and in $B\to K^*\mu^+\mu^-$, since the latter decay could be affected as well.
This mars the improvement due to the $B\to K$ lattice form factors mentioned above and it will be important to resolve this issue in the future.

\subsection{Confronting the data}\label{sec:data}

\begin{figure}[tp]
\centering
\includegraphics[width=0.49\textwidth]{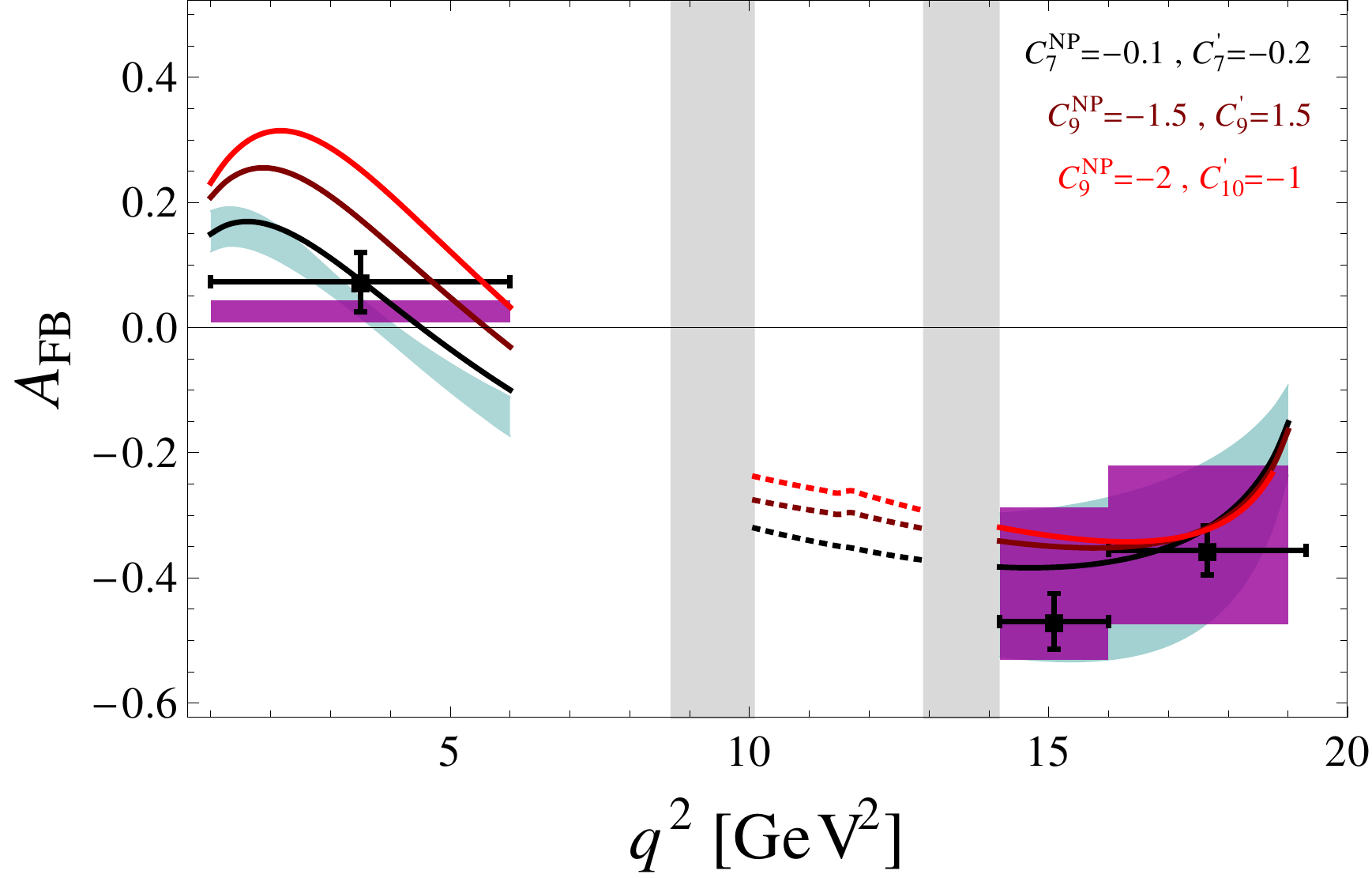} \includegraphics[width=0.49\textwidth]{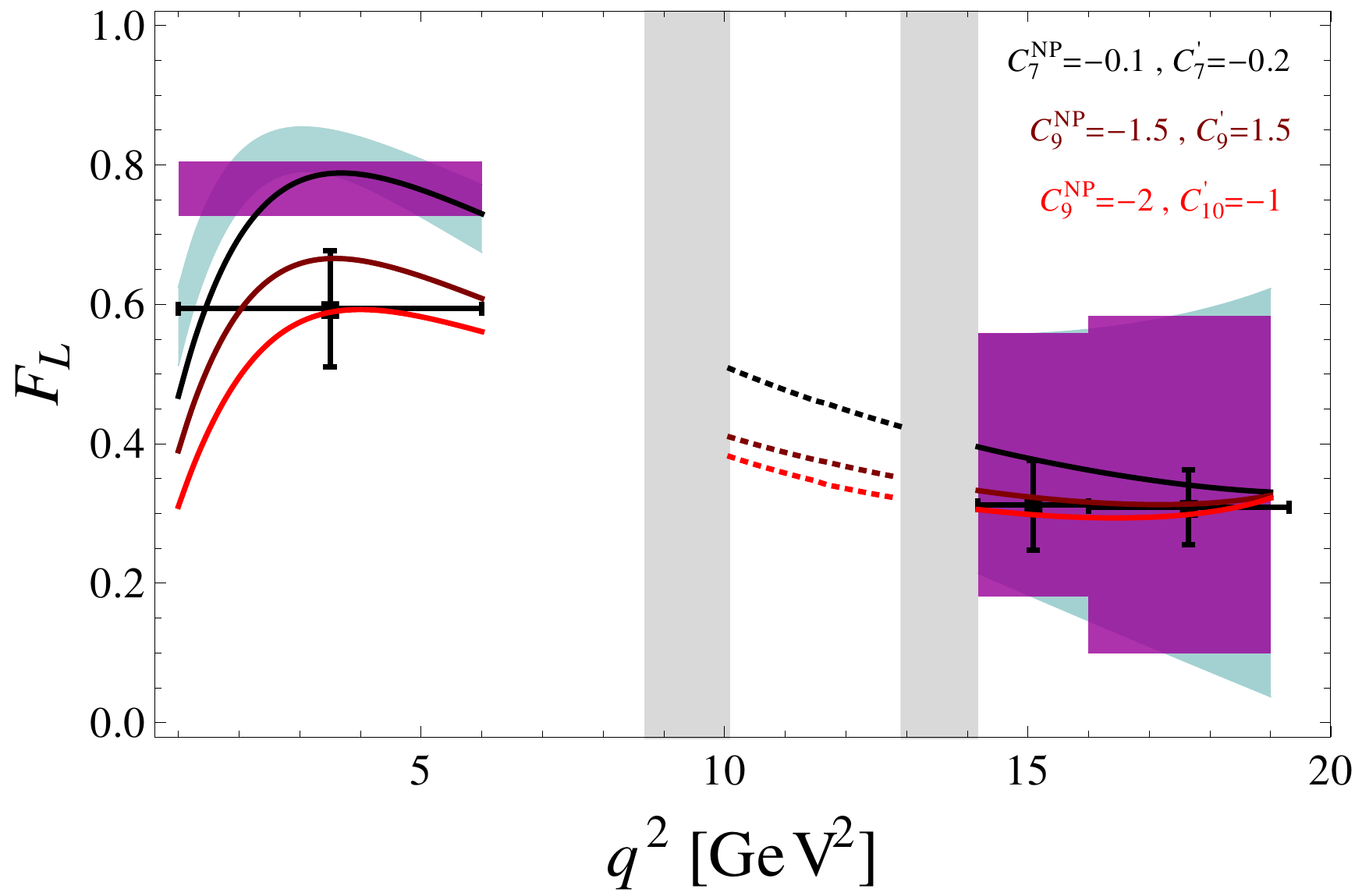} \\[16pt]
\includegraphics[width=0.49\textwidth]{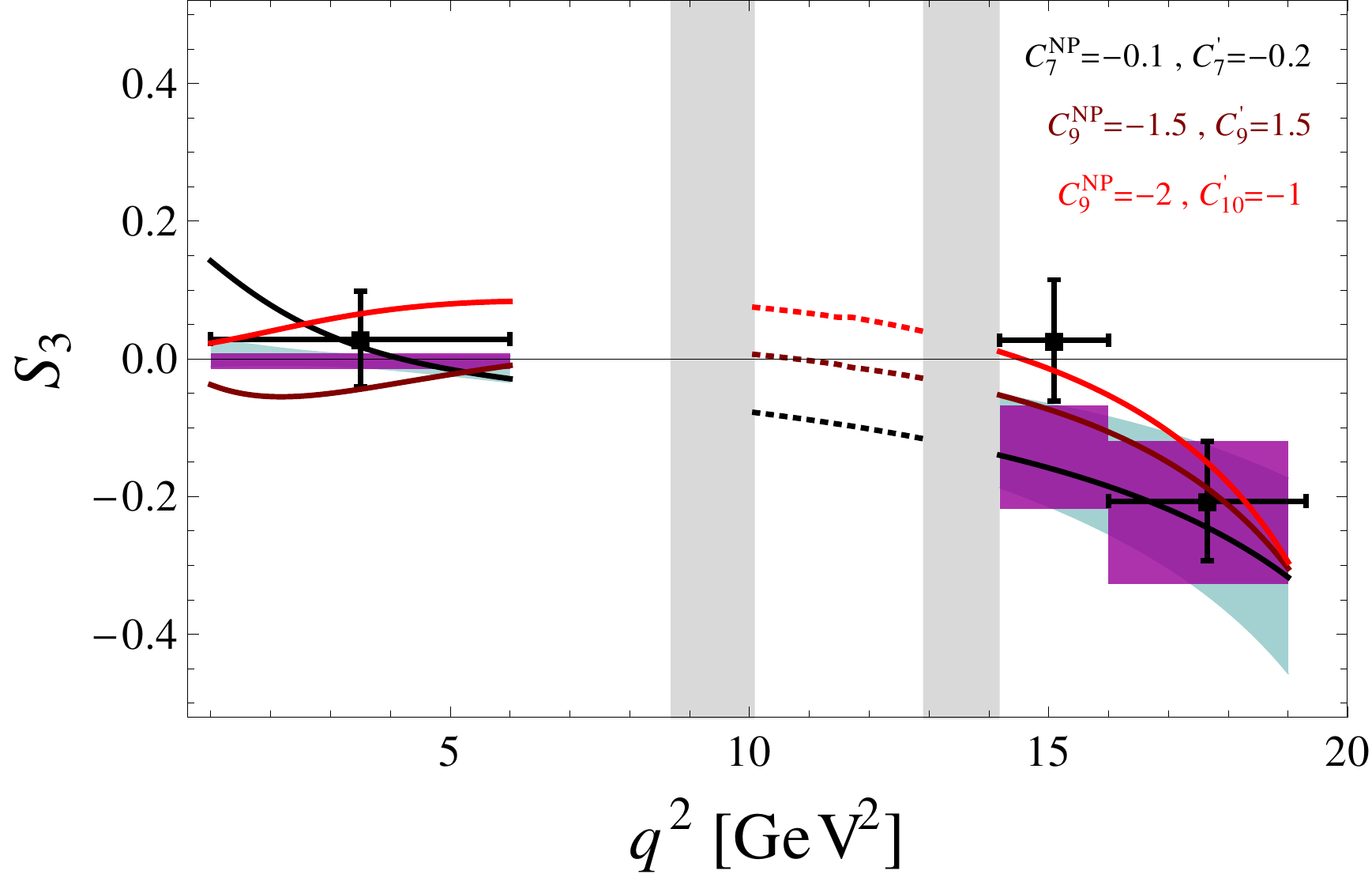} \includegraphics[width=0.49\textwidth]{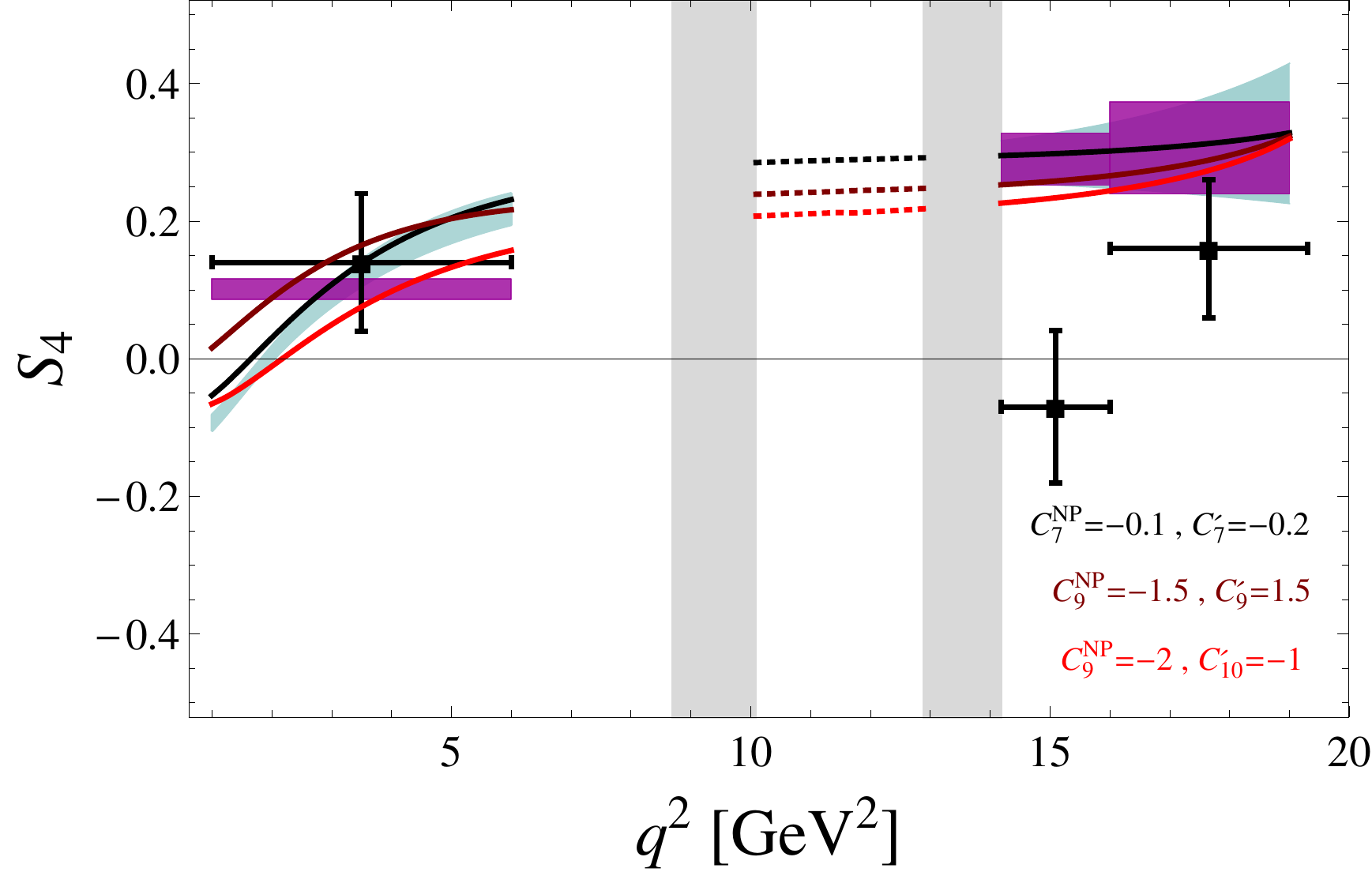} \\[16pt]
\includegraphics[width=0.49\textwidth]{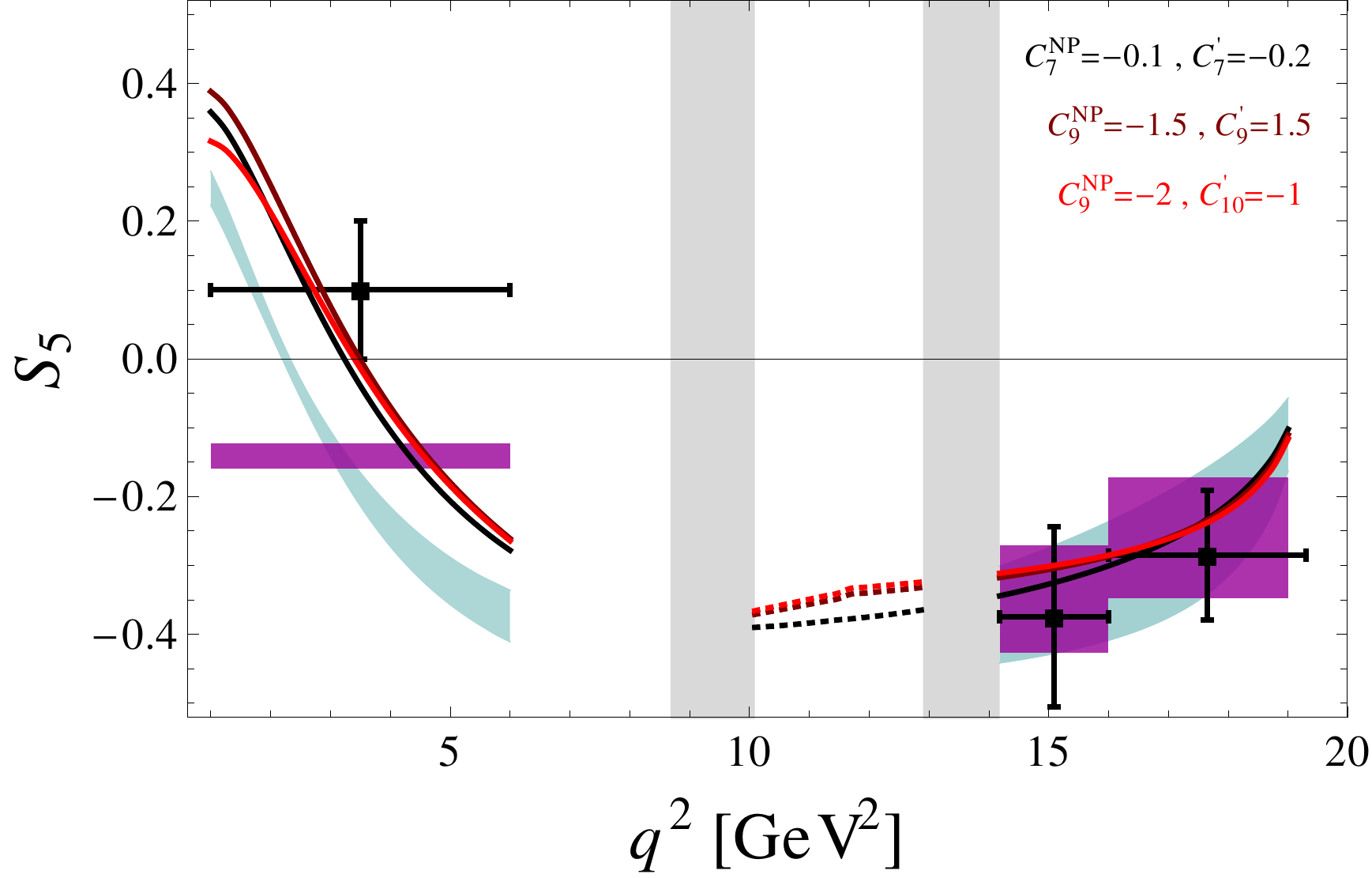}
\caption{The CP averaged angular observables $A_\text{FB}$, $F_L$, $S_3$, $S_4$, and $S_5$ as a function of the di-muon invariant mass squared $q^2$. Differential (binned) SM predictions are shown with light blue bands (purple boxes). The combined experimental data is represented by the black crosses. The black, brown, and red curves correspond to NP scenarios that reproduce the value of $S_5$ at low $q^2$ measured by LHCb.}
\label{fig:BKsll}
\end{figure}

Averaging all available data on $B\to K^*\mu^+\mu^-$, we can confront them with our SM predictions.
Figure~\ref{fig:BKsll} shows our
differential and binned SM predictions
(light blue bands and purple boxes) 
together with the combined experimental data (black crosses)
for the
observables $A_\text{FB}$, $F_L$, $S_3$, $S_4$, and $S_5$.
Our error estimates were described in detail in \cite{Altmannshofer:2012az}, the only change being the additional relative uncertainty at high $q^2$ as discussed in section~\ref{sec:data}.\footnote{%
We note that our treatment of form factors at high $q^2$ is based on the extrapolation of light-cone sum rule calculations \cite{Ball:2004rg} done in \cite{Bharucha:2010im}, which leads to particularly large uncertainties in $F_L$ compared to other approaches.}

Confronting theory and experiment, we observe three discrepancies:
\begin{itemize}
\item A deficit in $F_L$ at low $q^2$, mostly driven by the recent ATLAS measurement and to a lesser extent by BaBar data. As the ATLAS and BaBar data in the $[1,6]$ GeV$^2$ bin differ substantially from the LHCb, CMS, Belle and CDF results, we use the PDG averaging method to combine the data, i.e. we rescale the uncertainty of the weighted average by a factor of $\sqrt{\chi^2}$. Doing so, we still find a discrepancy with the SM prediction at a significance of $1.9\sigma$ in the $[1,6]$ GeV$^2$ bin (see appendix~\ref{app:combination} for more details).
\item A preference for an anomalously low value of $S_4$ at high $q^2$, measured only by LHCb. We find a significance of
$2.8\sigma$
in the $[14.18,16]$ GeV$^2$ bin.
\item A preference for an opposite sign in $S_5$ at low $q^2$, also measured only by LHCb. We find a significance of $2.4\sigma$ in the $[1,6]$ GeV$^2$ bin.\footnote{%
Experimental results for $S_4$ and $S_5$ in the $[1,6]$ GeV$^2$ region are not available yet. We therefore translate the results on $P_4^\prime$ and $P_5^\prime$ using the measured value of $F_L$ and get
$\langle S_4\rangle_{[1,6]} = 0.14 \pm 0.08$\,, $\langle S_5\rangle_{[1,6]} = 0.10 \pm 0.10$\,.%
}
\end{itemize}
We note that the significance of the tension in $S_5$ is substantially increased when considering only the $[4.3,8.68]$~GeV$^2$ bin~\cite{Aaij:2013qta}.
However, we consistently stick to the $[1,6]$~GeV$^2$ bin in the low-$q^2$ region to be conservative.

Using the alternative basis of observables suggested in \cite{Descotes-Genon:2013vna}, we obtain a tension of $2.1\sigma$ in $P_4'$ in the $[14.18,16]$~GeV$^2$ bin and of $2.6\sigma$ in $P_5'$ in the $[1,6]$~GeV$^2$ bin.
We stress that the increased significance in the case of $P_5'$ compared to $S_5$ is not due to a reduced theory uncertainty. In fact, using our error estimates, we obtain a relative uncertainty of $13\%$ in the low-$q^2$ bin in both cases. Rather, due to its normalization containing $F_L$, $P_5'$ feels both the tensions in $S_5$ and $F_L$ discussed separately above.
In the case of $P_4'$, the alternative normalization does lead to a reduced theory uncertainty with our choice of high-$q^2$ form factors.

We now turn to the discussion of new physics effects explaining the above tensions.

\subsection{Preliminary considerations}\label{sec:prelim}

Before turning to the numerical analysis, it is instructive to make some analytical considerations as to which Wilson coefficients have to be modified to explain the tensions in the data. An immediate observation is that all three tensions occur in CP-averaged observables, so there is no need to invoke non-standard CP violation, i.e.\ the Wilson coefficients can be kept real.

To get an analytical understanding of the dependence of the relevant observables on the Wilson coefficients, we can derive approximate expressions, valid for small NP contributions, neglecting interference terms between NP effects in different coefficients. We find\footnote{We stress the different sign of our definition of $S_4$ with respect to LHCb, see table~\ref{tab:dict}.}
\begin{align}
\langle F_L \rangle_{[1,6]}
&\simeq
+0.77
+0.25\,C_{7}^\text{NP}
\phantom{+0.00\,C_{7}'}\,\,\,
+0.05\,C_{9}^\text{NP}
-0.04\,C_{9}'
+0.04\,C_{10}' ~,
\\
\langle S_4 \rangle_{[14.18,16]}
&\simeq
+0.29
\phantom{+0.00\,C_{7}^\text{NP}}~
\phantom{+0.00\,C_{7}'}~
\phantom{+0.00\,C_{9}^\text{NP}}~~
-0.02\,C_{9}'
+0.03\,C_{10}' ~,
\\
\langle S_5 \rangle_{[1,6]}
&\simeq
-0.14
-0.59\,C_{7}^\text{NP}
-0.49\,C_{7}'
-0.09\,C_{9}^\text{NP}
-0.03\,C_{9}'
+0.10\,C_{10}' ~.
\end{align}

We can now qualitatively discuss the impact of NP effects in individual Wilson coefficients on the three anomalies.
\begin{itemize}
\item A negative NP contribution to $C_7$ would change both $F_L$ and $S_5$ in the right directions. Incidentally, this corresponds to constructive interference with the SM in $B\to X_s\gamma$, which is also slightly preferred by the data. $S_4$ at high $q^2$ would be unaffected.
\item A negative NP contribution to $C_9$ would also change both $F_L$ and $S_5$ in the right directions as well as leave $S_4$ unaffected.
\item A negative $C_7'$ could reduce the tension in $S_5$, but not the other two.
\item Non-zero $C_9'$ or $C_{10}'$ always worsen at least one of the tensions since they enter in all three observables with the same sign, while the experiment sees a deficit in $F_L$ and $S_4$ and a surplus in $S_5$.
\end{itemize}

Before quantifying this qualitative discussion by fitting the Wilson coefficients to all available data, we can already define some benchmark points that reconcile the theory with the experimental results for $B\to K^*\mu^+\mu^-$. In figure~\ref{fig:BKsll}, we show three NP scenarios in addition to the SM predictions and experimental data.
In the first scenario (black curves) we allow for sizable modifications of the dipole Wilson coefficients $C_7^\text{NP} = -0.1$ and $C_7^\prime = -0.2$, that are not excluded by the experimental data on $\text{BR}(B\to X_s \gamma)$ and $S_{K^*\gamma}$ (the time dependent CP asymmetry in $B \to K^*\gamma$). As can be seen from figure~\ref{fig:BKsll}, this scenario mainly affects $S_5$ but leaves all other observables, in particular $F_L$, approximately SM-like.  
In the second and third scenario, we use NP in $C_9$ to address the discrepancy in $S_5$. We set $C_9^\text{NP} = -1.5$, $C_9^\prime = 1.5$ (brown curves) and $C_9^\text{NP} = -2$, $C_{10}^\prime = -1$ (red curves). NP in $C_9^\prime$ and $C_{10}^\prime$ is switched on to avoid bounds from $\text{BR}(B\to K\mu^+\mu^-)$, as discussed in detail below. These scenarios also allow to accommodate a reduced $F_L$ compared to the SM. On the other hand, they also induce a slight tension with $A_\text{FB}$, as they increase the value of $A_\text{FB}$ and shift its zero crossing to higher $q^2$ values.

\subsection{Trying to reduce the tension in \texorpdfstring{$S_4$}{S4}}

\begin{figure}[tp]
 \centering
 \includegraphics[width=0.5\textwidth]{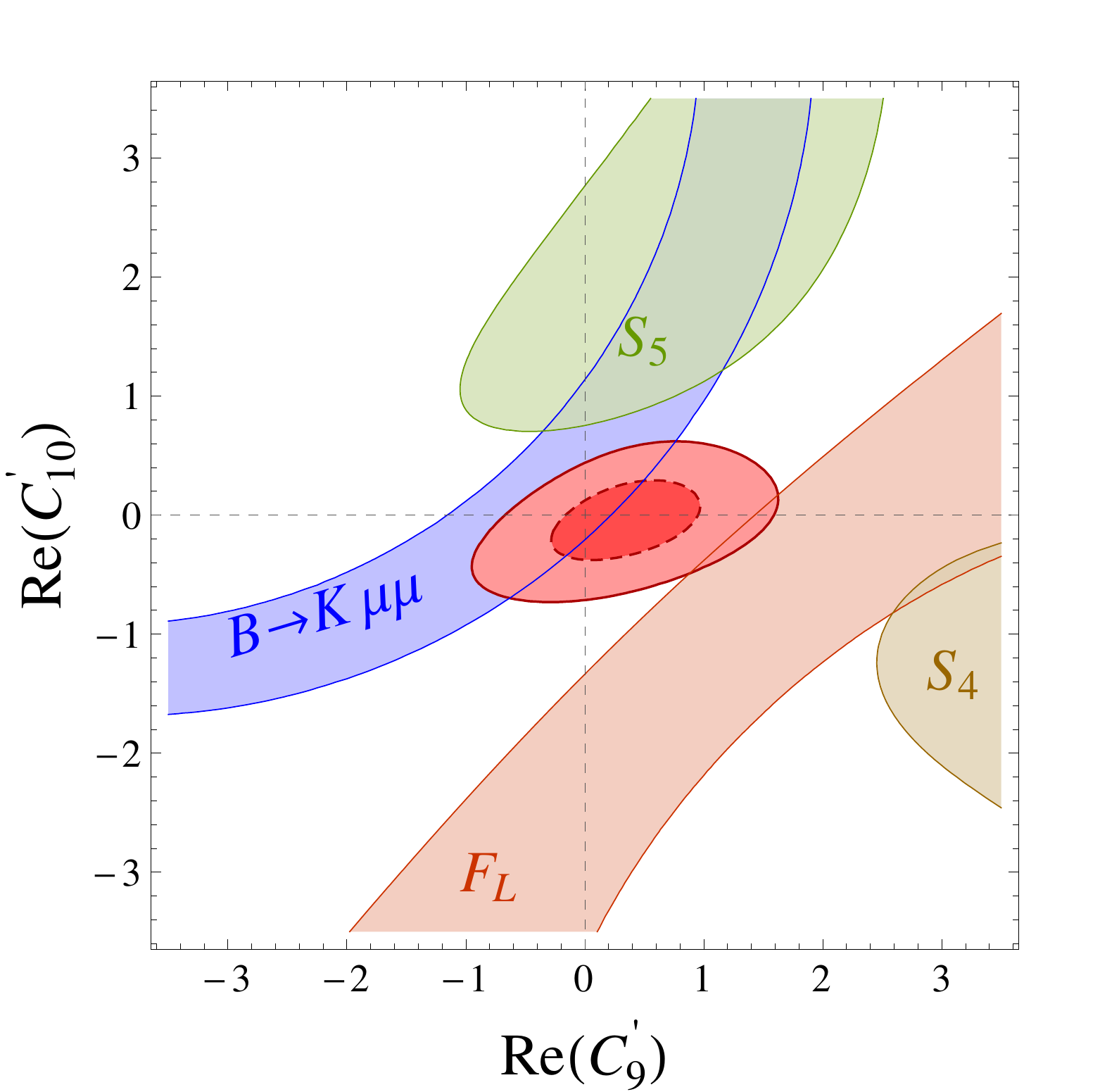}%
 \includegraphics[width=0.5\textwidth]{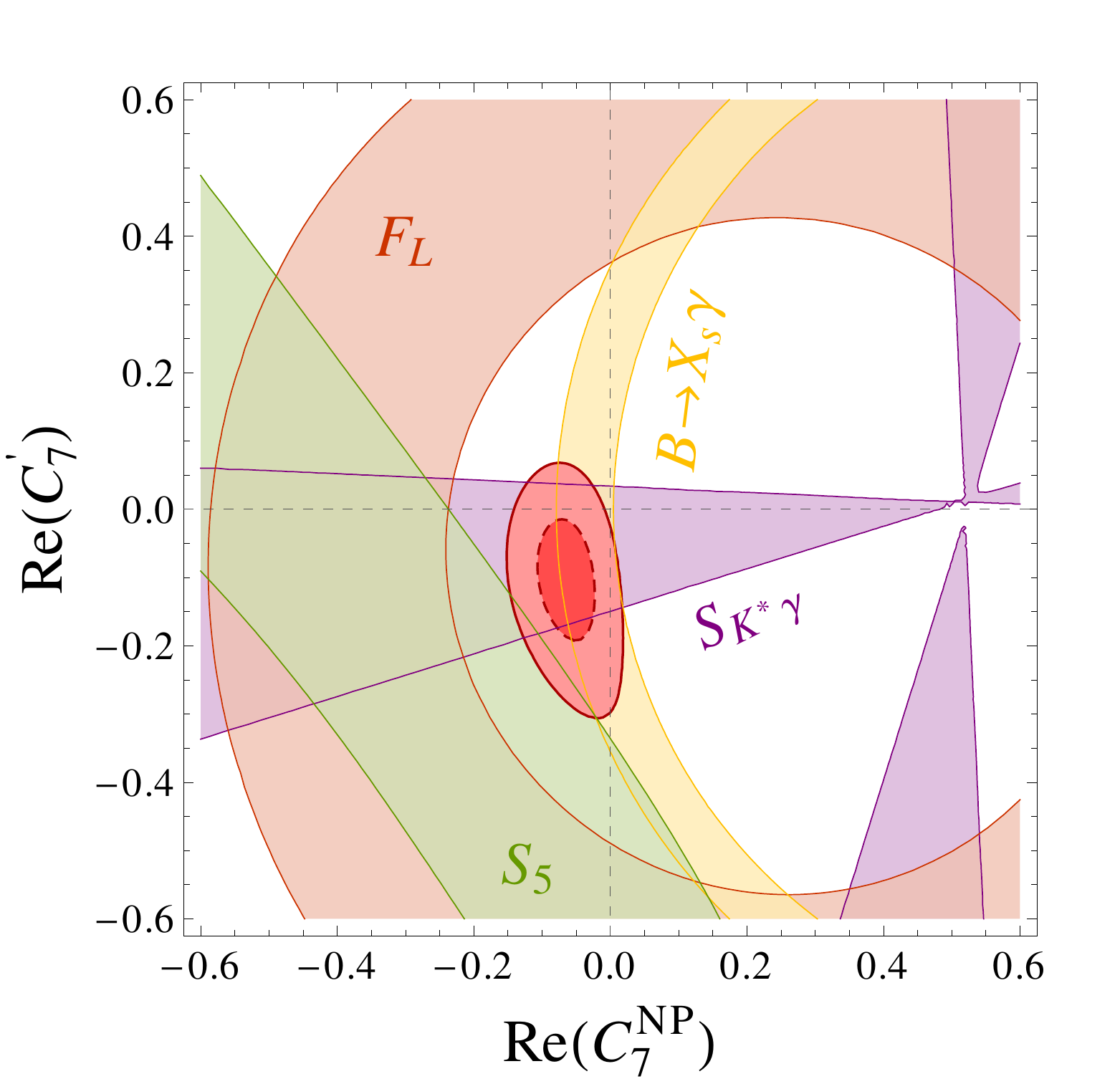}
 \caption{Constraints in the $C_9^\prime$-$C_{10}^\prime$ plane (left) and the  $C_7^\text{NP}$-$C_7^\prime$ plane (right). Combined $\Delta\chi=1,4$ contours are shown in red. Individual $\Delta\chi^2=1$ contours are also shown for $F_L$ (orange), $S_4$ (brown), $S_5$ (green), $\text{BR}(B\to K\mu^+\mu^-)$ (blue), $\text{BR}(B\to X_s \gamma)$ (yellow) and $S_{K^*\gamma}$ (purple).} 
 \label{fig:C9p10p77p}
\end{figure}

As discussed above, $S_4$ at high $q^2$ is mostly sensitive to $C_{9}'$ and $C_{10}'$, but NP effects in these coefficients would worsen either the tension in $F_L$ or the one in $S_5$.
To quantify this problem, in the left-hand panel of figure~\ref{fig:C9p10p77p}
we show the $\Delta \chi^2 = 1$ constraints on the $C_{9}'$-$C_{10}'$ plane from $S_4$, $S_5$, $F_L$ and the branching ratio of $B\to K\mu^+\mu^-$ as well as the combined $\Delta \chi^2 = 1$ and $\Delta \chi^2 = 4$  regions from all the constraints discussed in~\cite{Altmannshofer:2012az}.\footnote{We stress that in all the following plots, the red regions combine all relevant constraints, while the bands for individual observables are only shown for illustrative purposes for some of the most relevant observables. We also note that 68 and 95\% C.L. constraints on individual coefficients can be read off from the overlap of the bands with the axes, while the two-dimensional regions would correspond to 39 and 86\% C.L.}

Combining all constraints, the preferred region does not deviate significantly from the SM point $C_9'=C_{10}'=0$.
As expected, the tensions in $F_L$ and $S_5$ would require opposite NP effects. The tension in $S_4$ would require very large effects, which are anyway excluded. 
The $\text{BR}(B\to K\mu^+\mu^-)$ also gives an important constraint.
While compatible with the SM, also this observable is in tension with $F_L$ and $S_4$.

Another possibility would be to reduce the tension in $S_4$ by $C_9'$ or $C_{10}'$ and invoke NP contributions to the other Wilson coefficients to bring all constraints into agreement. However, we find that even varying all Wilson coefficients simultaneously and allowing for arbitrary CP phases, the tension in $S_4$ still remains at the $2\sigma$ level.
We thus conclude that the tension in $S_4$ at high $q^2$ cannot be explained by new physics with the operator basis~(\ref{eq:ops}).
In the following, we will therefore focus on NP effects that can reduce the tensions in $S_5$ and $F_L$ at low $q^2$.

\subsection{Effects in individual Wilson coefficients}

The considerations in section~\ref{sec:prelim} showed that if only a single Wilson coefficient is modified, the tensions in $F_L$ and $S_5$ can be reduced by negative NP effects in $C_7$ or $C_9$. 
The left-hand panel of figure~\ref{fig:C79} shows the $\Delta \chi^2 = 1$ constraints from $S_5$, $F_L$, $A_\text{FB}$, $\text{BR}(B\to K\mu^+\mu^-)$, and $\text{BR}(B\to X_s \gamma)$ in the $C_7^\text{NP}$-$C_9^\text{NP}$ plane.
Due to the strong constraint from $\text{BR}(B\to X_s \gamma)$, NP in the Wilson coefficient $C_7$ alone can improve the discrepancy in the $S_5$ data only slightly.
With NP in $C_9$, the improvement is better; however both $B\to K\mu^+\mu^-$ and $A_\text{FB}$ limit the size of the NP contributions, so that the tensions cannot be solved fully.
The best-fit values for effects in $C_7^\text{NP}$ and $C_9^\text{NP}$ individually are listed in the first two rows of table~\ref{tab:scen} as NP scenarios $(7)$ and $(9)$.

\subsection{Effects in two Wilson coefficients}

We now consider simultaneous NP effects in two Wilson coefficients.
We start with scenarios without NP contributions to $C_9$ or $C_9'$, as this is the case in two of the three concrete NP scenarios discussed in section~\ref{sec:NPmodels}.
In this case, an improvement of the tensions comparable to the case with $C_9$ only
is possible by small, negative contributions to both $C_7$ and $C_7'$.
As shown in the right-hand panel of figure~\ref{fig:C9p10p77p}, in this scenario $(77')$ the tensions in $F_L$ and $S_5$ can be reduced, but the size of the effects is limited by the constraints from $\text{BR}(B\to X_s \gamma)$ and $S_{K^*\gamma}$.

\begin{figure}[tp]
 \centering
 \includegraphics[width=0.5\textwidth]{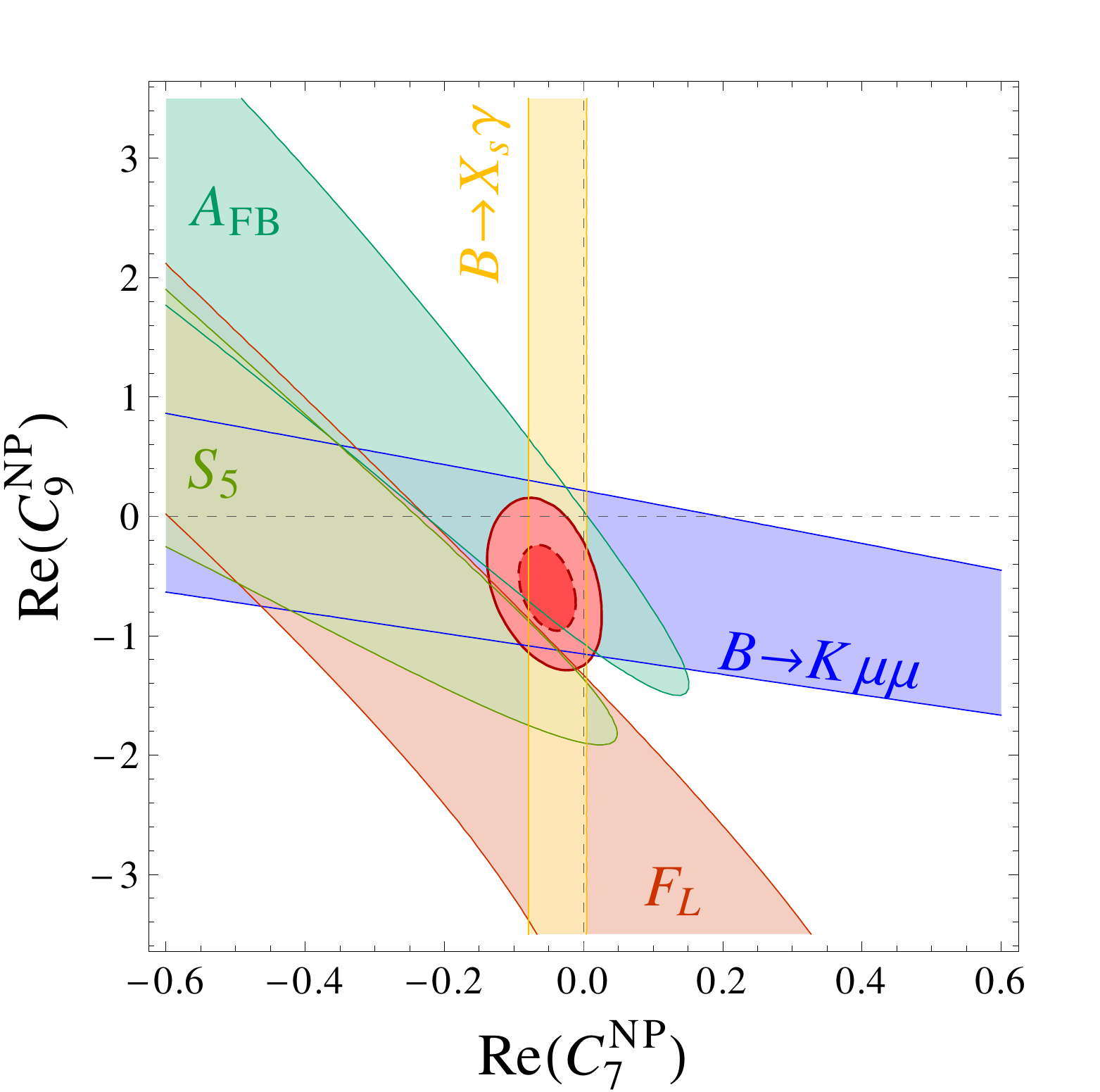}%
 \includegraphics[width=0.5\textwidth]{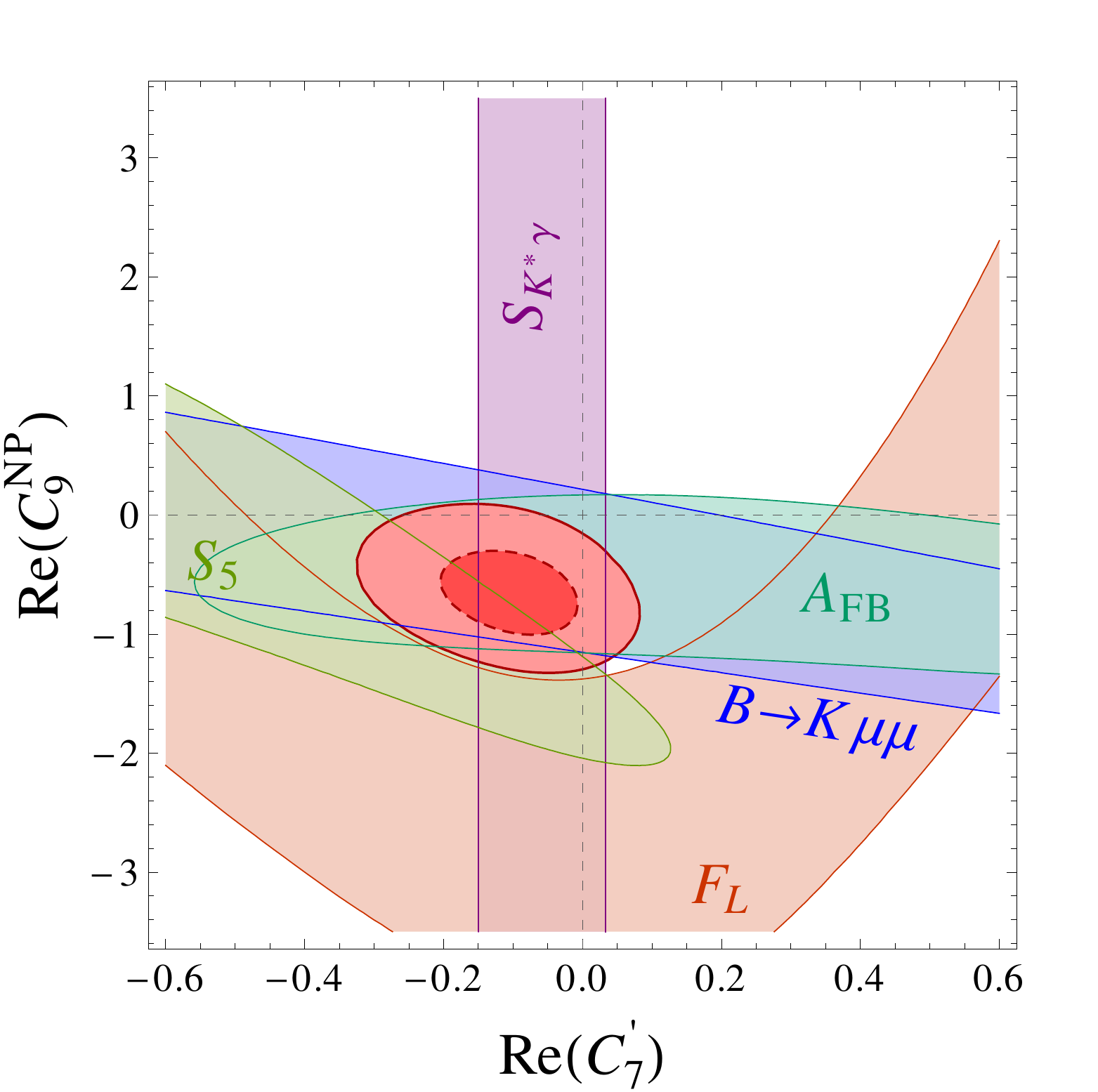}%
 \caption{Constraints in the $C_7^\text{NP}$-$C_9^\text{NP}$ plane (left) and the $C_7^\prime$-$C_9^\text{NP}$ plane (right).
 }
\label{fig:C79}
\end{figure}

\begin{figure}[tp]
 \centering
 \includegraphics[width=0.5\textwidth]{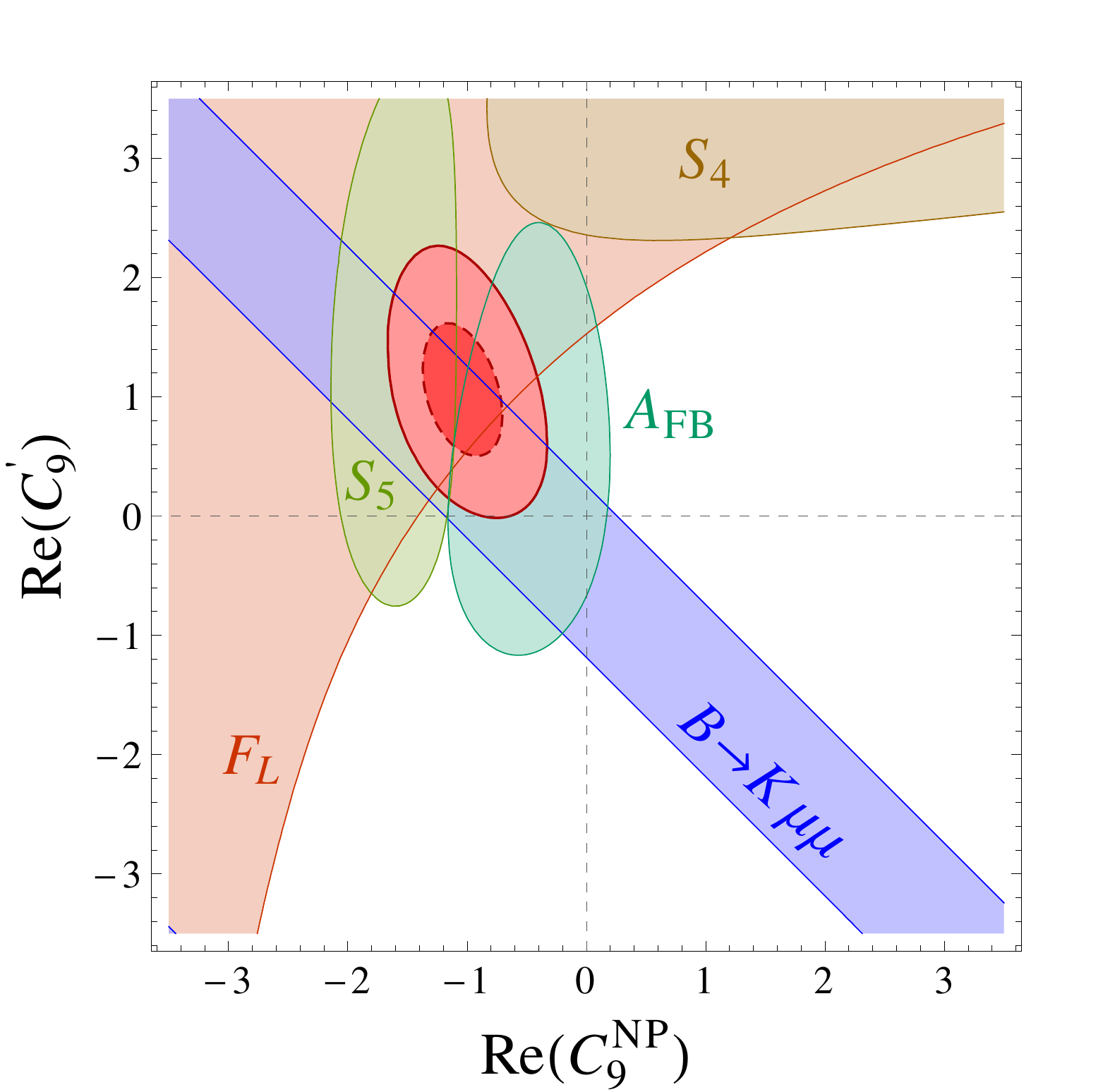}%
 \includegraphics[width=0.5\textwidth]{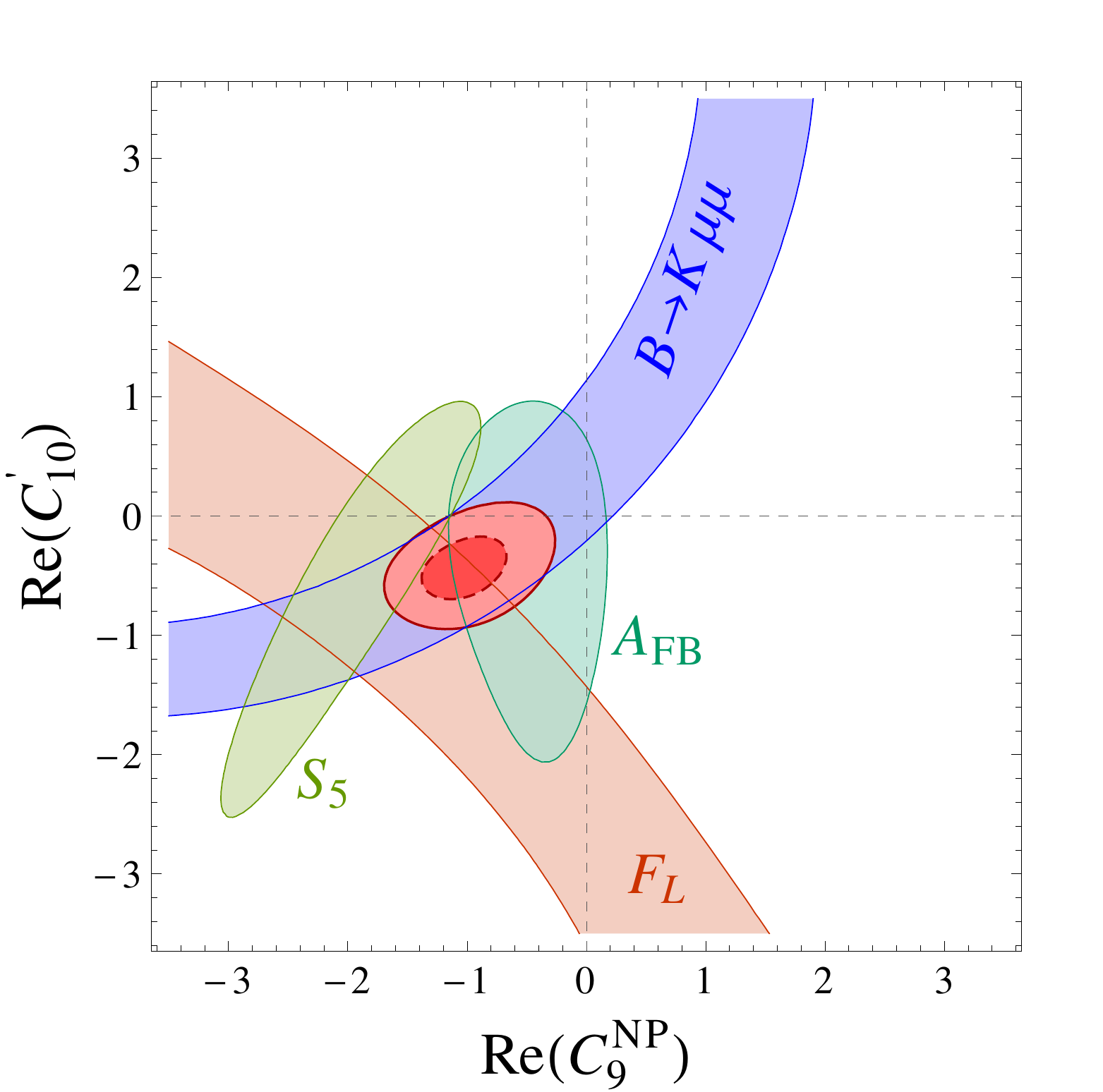}%
 \caption{Constraints in the $C_9^\text{NP}$-$C_9^\prime$ plane (left) and the $C_9^\text{NP}$-$C_{10}^\prime$ plane (right).
 }
 \label{fig:C910}
\end{figure}

A slightly better fit can be obtained in the presence of NP contributions to $C_9$, in combination with $C_7$ or $C_7'$.
In these scenarios, denoted 
$(97)$ and $(97')$ and shown in figure~\ref{fig:C79}, the situation is similar to the $(77')$ case:
the tensions in $S_5$ and $F_L$ can only be reduced partly since
various observables, which are in good agreement with the SM, limit the size of the NP contributions.
In particular,
NP in $C_9$ or $C_9'$ is limited by $A_\text{FB}$ and BR($B\to K\mu\mu$).
The fits thus find a compromise between improving the agreement in $S_5$ and $F_L$, but worsening it in other observables.

In fact, the best two-coefficient fits are obtained when combining  $C_9$ with $C_9'$ or $C_{10}'$, scenarios $(99')$ and $(910')$ shown in figure~\ref{fig:C910}.
Here the tensions in $S_5$ and $F_L$ can be almost completely removed, while the constraints from radiative decays are absent of course and the constraint from BR($B\to K\mu\mu$) is avoided by simultaneous contributions to $C_9'$ with same sign or to $C_{10}'$ with opposite sign, leading to a cancellation.
Remaining is only a slight tension between $S_5$ and $F_L$ on the one side and $A_\text{FB}$ on the other side.

\begin{figure}[tp]
\centering
\includegraphics[width=0.75\textwidth]{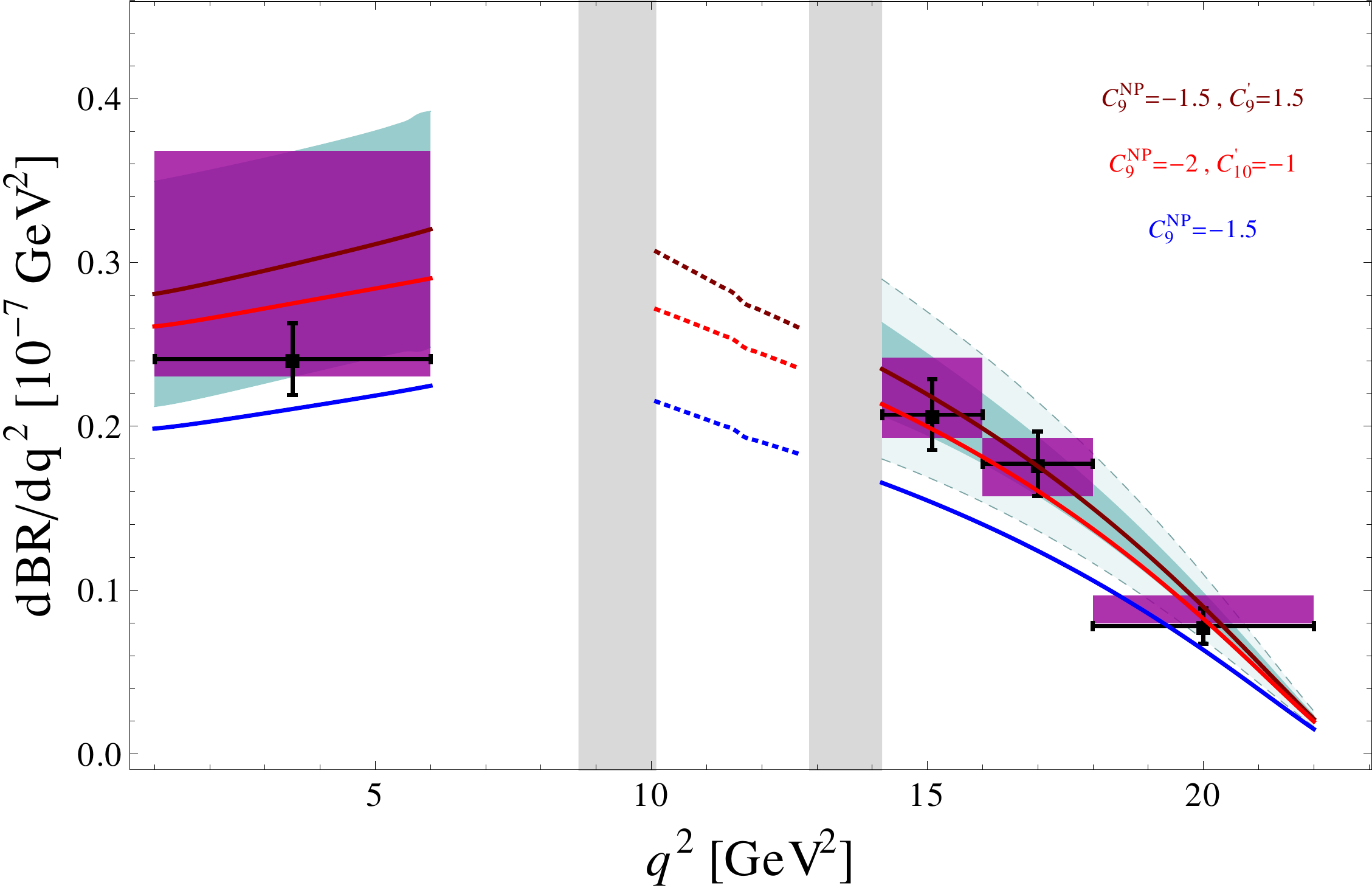}
\caption{The branching ratio of $B^+\to K^+\mu^+\mu^-$ as a function of the di-muon invariant mass squared $q^2$. Differential (binned) SM predictions are shown with light blue bands (purple boxes). The experimental data is represented by the black crosses. The blue, brown, and red curves correspond to NP scenarios that reproduce the value of $S_5$ at low $q^2$ measured by LHCb. At high $q^2$, the darker shaded region and the purple boxes show the error estimates with lattice form factors before adding an additional 20\% relative uncertainty to account for the effect of charmonium resonances. Our full uncertainty is shown by the lighter shaded region with dashed borders.}
\label{fig:BKll}
\end{figure}

To understand this cancellation better, it is instructive to have a closer look at the branching ratio of $B\to K\mu^+\mu^-$. Since the axial vector current does not have a $B\to K$ matrix element, the decay is only sensitive to the sum of primed and unprimed Wilson coefficients. Due to the new lattice form factors~\cite{Bouchard:2013mia}, the constraint from the high-$q^2$ branching ratio is particularly important, even adding the 20\% relative uncertainty due to possible resonance contributions discussed in section~\ref{sec:fit}. Expanding in small NP contributions, one finds\footnote{We consider the branching ratio in the entire high-$q^2$ region for indicative purposes. In the numerical analysis, we use three separate bins, as in the LHCb analysis.}
\begin{multline}
10^{7}\times \text{BR}(B^+\to K^+\mu^+\mu^-)_{[14.18,22]}
\simeq
1.11+
\\
+0.22\,(C_{7}^\text{NP}+C_{7}')
+0.27\,(C_{9}^\text{NP}+C_{9}')
-0.27\,(C_{10}^\text{NP}+C_{10}')
\,,
\label{eq:BKmm}
\end{multline}
where LHCb has measured
\begin{align}
10^{7}\times\text{BR}(B^+\to K^+\mu^+\mu^-)_{[14.18,22]}
&=
1.04\pm0.12 \,.
\end{align}
For the relative uncertainty of the SM prediction, we find 10\% {\em before} adding the additional relative 20\% uncertainty, in agreement with \cite{Bouchard:2013mia}.
Since the central values fit perfectly, any NP contribution is strongly constrained. However, it is always possible to cancel the NP contribution in an unprimed coefficient by a corresponding contribution to the primed one, or to cancel contributions to $C_9^{(\prime)}$ by corresponding contributions to $C_{10}^{(\prime)}$.
This is illustrated in figure~\ref{fig:BKll}, showing the SM prediction for the differential branching ratio. The brown and red curves are the same NP scenarios as in figure~\ref{fig:BKsll} and one can see the effect of the cancellation between $C_9$ and $C_9'$ or $C_{10}'$. In a scenario with NP in $C_9$ only, that would be enough to solve the tensions in $F_L$ and $S_5$, one would get a significant reduction of the branching ratio, as shown by the blue curve, which is disfavoured even with our conservative error estimates.

We conclude this section by considering the best fit that can be obtained by varying {\em all} Wilson coefficients simultaneously. If the CP phase is assumed to be aligned with the SM, the best-fit point, shown in the penultimate row of  table~\ref{tab:scen}, is a combination of the preferred values in the two-coefficient fits and prefers small $C_{10}$ and $C_{10}'$. However, it is only slightly better than scenario $(99')$.
Allowing completely arbitrary CP phases, another slight reduction of the $\chi^2$ is possible (at the cost of a large number of free parameters), as shown in the last row of  table~\ref{tab:scen}.

\begin{table}[tp]
\renewcommand{\arraystretch}{1.4}
\centering
\begin{tabular}{lccccccr}
\hline\hline
\bf\boldmath Scenario &$C_7^\text{NP}$ & $C_7'$ & $C_9^\text{NP}$ & $C_9'$ & $C_{10}^\text{NP}$ & $C_{10}'$ & $\Delta \chi^2(\text{SM})$\\
\hline
\bf\boldmath (7) & $-0.07\scriptstyle\pm0.04$ &&&&&& $3.4$ \\
\bf\boldmath (9) &&& $-0.8\scriptstyle\pm0.3$ &&&& $\mathbf{4.3}$ \\
\hline
\bf\boldmath($77'$) & $-0.06\scriptstyle\pm0.04$ & $-0.1\scriptstyle\pm0.1$ &&&&& $4.7$ \\
\bf\boldmath($97$) & $-0.05\scriptstyle\pm0.04$ && $-0.6\scriptstyle\pm0.3$ &&&& $6.0$ \\
\bf\boldmath($97'$) && $-0.1\scriptstyle\pm0.1$ & $-0.7\scriptstyle\pm0.3$ &&&& $5.5$ \\
\bf\boldmath($99'$) &&& $-1.0\scriptstyle\pm0.3$ & $+1.0\scriptstyle\pm0.5$ &&& $\mathbf{8.3}$ \\
\bf\boldmath($910'$) &&& $-1.0\scriptstyle\pm0.3$ &&& $-0.4\scriptstyle\pm0.2$ & $7.0$ \\
\hline
\bf\boldmath Real & $-0.03$ & $-0.11$ & $-0.9$ & $+0.7$ & $-0.1$ & $-0.2$ & $\mathbf{10.8}$ \\
\bf\boldmath Complex & $\begin{smallmatrix}+0.03\\~~~+0.09i\end{smallmatrix}$ & $\begin{smallmatrix}-0.23\\~~~-0.23i\end{smallmatrix}$ & $\begin{smallmatrix}-1.9\\~~~+1.2i\end{smallmatrix}$ & $\begin{smallmatrix}+1.2\\~~~+3.3i\end{smallmatrix}$ & $\begin{smallmatrix}+1.6\\~~~-0.1i\end{smallmatrix}$ & $\begin{smallmatrix}+1.0\\~~~+1.6i\end{smallmatrix}$ & $\mathbf{14.1}$ \\
\hline\hline
\end{tabular}
\caption{Best-fit values and reduction in the total $\chi^2$ with respect to the SM fit when varying individual Wilson coefficients, pairs of coefficients or all coefficients simultaneously. In all but the last case, we assume the coefficients to be real. We omit cases where $\Delta \chi^2(\text{SM})<4$ (except the $C_7$ case) or where one of the coefficients has a best-fit value of 0.
For the one- and two-coefficient fits, we also give the $1\sigma$ fit ranges.
}
\label{tab:scen}
\end{table}

\section{Implications for models of new physics}\label{sec:NPmodels}

The preferred values of the individual Wilson coefficients that address the observed tensions in $B \to K^* \mu^+\mu^-$ can be translated into NP scales in a model independent way by defining NP effects to the effective Hamiltonian as $\Delta \mathcal H_\text{eff}=-\mathcal O_i/\Lambda_i^2$.
In the case of the operators $\mathcal O_9^{(\prime)}$ and $\mathcal O_7^{(\prime)}$ we find
\begin{equation}
 \Lambda_{9^{(\prime)}} \simeq (35~\text{TeV}) \left( \frac{1.0}{|C_9^{(\prime)}|} \right)^{1/2} ~,~~ \Lambda_{7^{(\prime)}} \simeq (90~\text{TeV}) \left( \frac{0.1}{|C_7^{(\prime)}|} \right)^{1/2} ~,
\end{equation}
where we explicitly factored out interesting values for the NP contributions to the Wilson coefficients.

Typically, one expects dipole operators to arise only at the loop level. In various concrete models of NP, like the MSSM, also the semileptonic operators are induced only at 1-loop. We therefore repeat the above exercise, including an explicit loop factor $1/(4\pi)^2$ in the effective Hamiltonian, leading to
\begin{equation}
 \Lambda_{9^{(\prime)}}^\text{loop} \simeq (2.8~\text{TeV}) \left( \frac{1.0}{|C_9^{(\prime)}|} \right)^{1/2} ~,~~ \Lambda_{7^{(\prime)}}^\text{loop} \simeq (7.5~\text{TeV}) \left( \frac{0.1}{|C_7^{(\prime)}|} \right)^{1/2} ~.
\end{equation}
We learn that even in the case of loop suppression, the observed discrepancies can be explained by very heavy NP, at the border or outside the direct reach of the LHC. Note however, that such NP is required to have maximal, i.e. $O(1)$, mixing between the bottom and strange flavour as well as $O(1)$ couplings to SM leptons.
In models with Minimal Flavour Violation, where the $b\to s$ transition is suppressed by the same CKM factors as in the SM, the scales are another factor of 5 smaller.

We now discuss to which extent these generic expectations are modified in concrete models of NP.
Our focus is on well-motivated extensions of the SM, like the Minimal Supersymmetric Standard Model (MSSM) or models with partial compositeness.
However, as we will describe in detail in sections~\ref{sec:MSSM} and \ref{sec:PC}, sizable NP contributions to $C_9$ or $C_9'$ are not expected in these models. We thus start our discussion with the introduction of a heavy, neutral, flavour-changing gauge boson that could generate such effects.

\subsection{Flavour-changing neutral gauge boson}

An obvious way to generate NP contributions to $C_9^{(\prime)}$, as preferred by the fit to the $B \to K^* \mu^+\mu^-$ data, is through tree-level exchange of a
heavy neutral gauge boson, i.e.\ a $Z^\prime$, with a flavour-changing $b \to s$ couplings.
Tree-level flavour changing couplings of a $Z^\prime$ can arise for example in $U(1)^\prime$ models with family non-universal charges~\cite{Langacker:2000ju}, in the ``little flavor'' model~\cite{Sun:2013cza}, and also in the ``effective $Z^\prime$'' setup~\cite{Fox:2011qd}, where the $Z^\prime$ couples to SM particles only through higher dimensional operators. The latter framework allows to treat couplings of the $Z^\prime$ to SM fermions essentially as free parameters.  
For recent works on the effects of flavour-changing $Z^\prime$'s in $B$ physics see~\cite{Barger:2009qs,Dighe:2012df,Buras:2012jb} and references therein.

We parameterize the relevant couplings of the $Z^\prime$ to bottom and strange quarks as well as muons in the following way
\begin{equation} \label{Zprime_couplings}
\mathcal L \supset \frac{g_2}{2c_W} \Big[ \bar s \gamma^\mu (g^L_{bs} P_L + g^R_{bs} P_R) b + \bar\mu \gamma^\mu (g^V_\mu + \gamma_5 g^A_\mu) \mu \Big] Z^\prime_\mu ~,
\end{equation}
where we kept both vector and axial-vector coupling to muons for completeness. 
Integrating out the $Z^\prime$ leads to 
\begin{equation}
\frac{e^2}{16\pi^2} (V_{ts}^*V_{tb}) \Big\{ C_9^\text{NP}, C_9^\prime, C_{10}^\text{NP} , C_{10}^\prime \Big\} = \frac{m_Z^2}{2 m_{Z^\prime}^2} \Big\{ g^L_{bs} g_\mu^V , g^R_{bs} g_\mu^V , g^L_{bs} g_\mu^A , g^R_{bs} g_\mu^A \Big\} ~.
\label{eq:Zp}
\end{equation}
As the $Z^\prime$ contribution is not loop suppressed and the flavour-changing couplings $g_{bs}^{L,R}$ can in principle be of order 1, very heavy $Z^\prime$ can accommodate a $C_9^\text{NP} \sim -1$.
It is interesting to note from eq.~(\ref{eq:Zp}) that the scenarios (9) and ($99'$) of table~\ref{tab:scen} can be realized with a single $Z'$, but not scenario $(910')$, since one always has $C_9^\text{NP}C_{10}'=C_{10}^\text{NP}C_9'$.

In addition to the contributions to the $b \to s \mu\mu$ transition, the flavour-changing $Z^\prime$ will necessarily induce also effects in $B_s$ mixing at the tree level. The effects in the $B_s$ mass difference $\Delta M_s$ can be written as
\begin{equation} \label{eq:DMs}
\frac{\Delta M_s}{\Delta M_s^\text{SM}}
\simeq 1 + \frac{m_Z^2}{m_{Z^\prime}^2} \Big[ (g_{bs}^L)^2 + (g_{bs}^R)^2 - 9.7 (g_{bs}^L)(g_{bs}^R) \Big] \left( \frac{g_2^2}{16\pi^2} (V_{ts}^*V_{tb})^2 S_0 \right)^{-1}~,
\end{equation}
where the SM loop function is $S_0 \simeq 2.3$. To arrive at~(\ref{eq:DMs}), we took the matching scale to be $1$~TeV and used the hadronic matrix elements collected in~\cite{Buras:2012jb}.
The contributions are particularly large if both left- and right-handed couplings are present, due to the larger hadronic matrix elements of the generated $B_s$ mixing operators with a left-right chirality structure.
From the experimental side, the mass difference $\Delta M_s$ is measured with very high precision~\cite{Abulencia:2006mq,Aaij:2013mpa}. Uncertainties in the SM prediction, stemming mainly from the limited precision of the hadronic matrix elements and CKM factors, allow for $O(10\%)$ NP contributions to this observable. For a given $Z^\prime$ mass, this limits the allowed size of the flavour-changing couplings $g_{bs}^{L,R}$ significantly.

\begin{figure}[tp]
\centering
\includegraphics[width=0.55\textwidth]{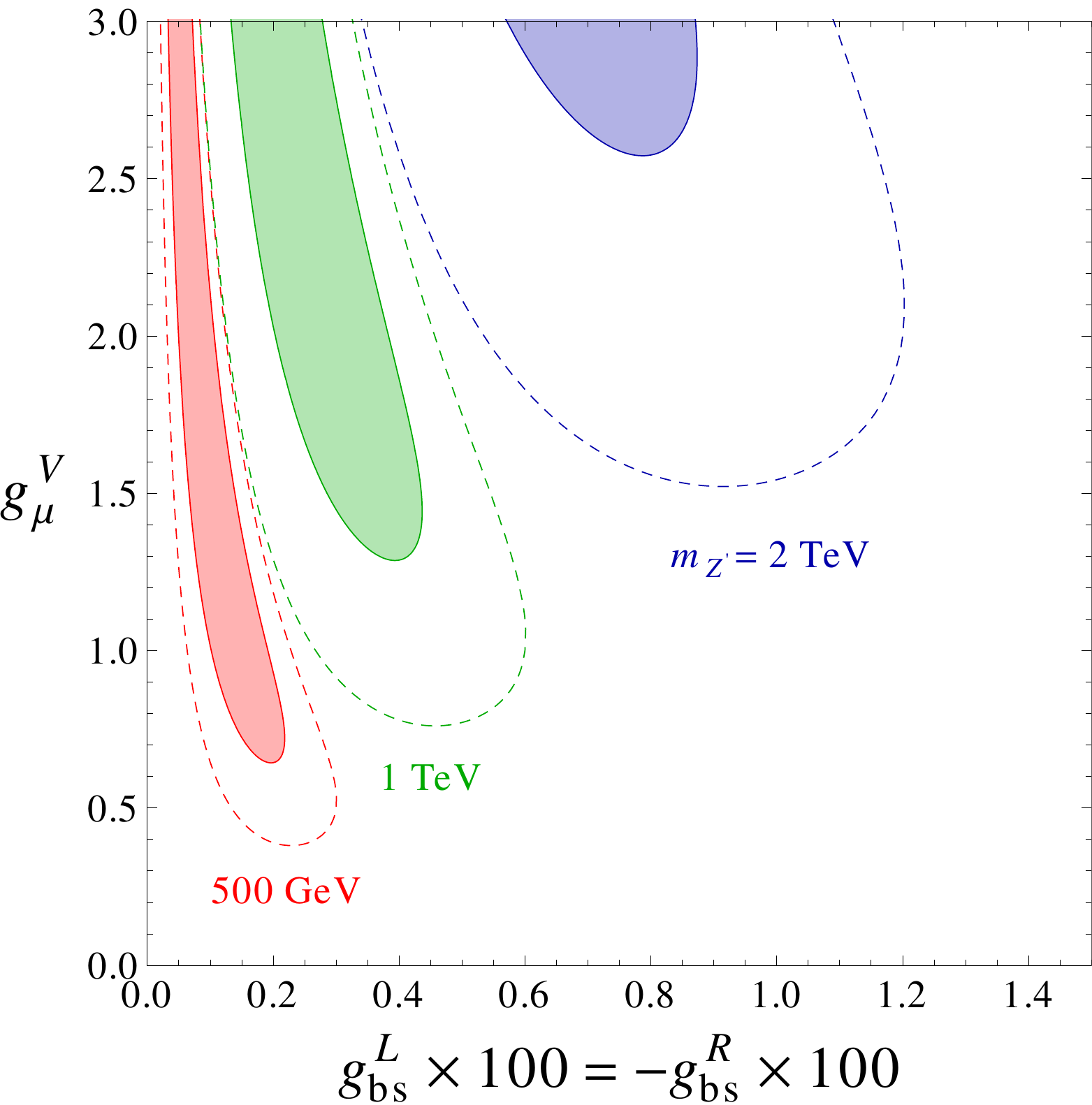}
\caption{Preferred values for the $Z^\prime$ vector couplings to muons ($g_\mu^V$) and $Z^\prime$ flavour-changing $b\to s$ couplings ($g_{bs}^{L,R}$), taking into account the best fit values for $C_9$ and $C_9^\prime$ as well as constraints from $B_s$ mixing. We set $g_{bs}^R = -g_{bs}^L$. The red/green/blue regions correspond to $Z^\prime$ masses of $m_{Z^\prime} = 500/1000/2000$~GeV. The solid (dashed) contours indicate $1\sigma$ and $2\sigma$ regions, respectively. Note the normalization of the $Z^\prime$ couplings in eq.~(\ref{Zprime_couplings}).}
\label{fig:Zprime}
\end{figure}

In figure~\ref{fig:Zprime} we show the preferred regions in the plane of the flavour-changing couplings $g_{bs}^{L,R}$ and the muon coupling $g_\mu^V$ for fixed values of the $Z^\prime$ mass of $m_{Z^\prime} = 500$~GeV (red), 1~TeV (green), and 2~TeV (blue).
We set $g_{bs}^L=-g_{bs}^R$ to reproduce scenario $(99')$ in table~\ref{tab:scen}
and take into account the best fit regions for the Wilson coefficients $C_9^{(\prime)}$ as well as the constraint from $\Delta M_s$. 
For a fixed value of the $Z^\prime$ coupling to muons, an explanation of the $B \to K^* \mu^+\mu^-$ anomalies at the 1$\sigma$ level, together with the constraint from $B_s$ mixing imply an upper bound on the $Z^\prime$ mass. 
We find $m_{Z^\prime} \lesssim g_\mu^V \times 780$~GeV if $g_{bs}^L = -g_{bs}^R$.

There exist strong bounds on $Z^\prime$ bosons from collider searches for di-jet and di-lepton resonances~\cite{CMS:kxa,CMS-PAS-EXO-12-061,ATLAS:2013jma}. A sequential $Z^\prime$, i.e. a $Z^\prime$ with the same couplings to SM fermions as the SM $Z$, is excluded up to masses of $\sim 3$~TeV. A lighter $Z^\prime$ can be viable if its production cross section is suppressed or its branching ratios into SM fermions is smaller. As the $Z^\prime$ considered here has to have sizable couplings to muons in order to account for the $B \to K^* \mu^+\mu^-$ anomalies, it is very plausible that its branching ratio to light leptons is at least of the same order as the corresponding branching ratio of the SM $Z$. We then find that in order to avoid the bounds from di-lepton resonance searches, the considered $Z^\prime$ with a mass $\lesssim 1$~TeV has to have couplings to first generation quarks that are suppressed by at least one order of magnitude compared to the corresponding couplings of the SM $Z$. 

As the $Z^\prime$ considered here couples dominantly to the charged lepton vector current, $SU(2)_L$ invariance necessarily implies also a coupling of the $Z^\prime$ to neutrinos. Therefore, generically, the $Z^\prime$ will also lead to NP effects in the neutrino modes $B \to K^{(*)} \nu\bar\nu$.
To estimate the size of the expected effects, we neglect possible $SU(2)_L$ breaking corrections and work in the limit\footnote{Note that, given the present uncertainties in the $B \to K^* \mu^+\mu^-$ data, a sizable axial-vector coupling to leptons cannot be excluded yet. Allowing for non-zero axial-vector couplings to leptons, could either enhance or suppress couplings to neutrinos.}
\begin{equation}
 g_\mu^A = g_\mu^R - g_\mu^L = 0~,~~~ g_\nu^L = g_\mu^L = \frac{1}{2} g_\mu^V ~.
\end{equation}
Using the results in~\cite{Altmannshofer:2009ma}, we find that $C_9^\text{NP} \simeq -1$ implies a slight enhancement of the branching ratios BR$(B \to K^* \nu\bar\nu)$ and BR$(B \to K \nu\bar\nu)$ of the order of 15\%. It will be challenging to reach such a precision at Belle II~\cite{Aushev:2010bq}.

\subsection{The Minimal Supersymmetric Standard Model} \label{sec:MSSM}

It is well known, that in the MSSM, large NP contributions to the dipole Wilson coefficients $C_7$ and $C_7^\prime$ can arise easily in large regions of parameter space. On the other hand, as we will show explicitly, the vector coefficients $C_9$ and $C_9^\prime$ remain to a good approximation SM-like throughout the viable MSSM parameter space, even if we allow for completely generic flavour mixing in the squark sector.

\begin{figure}[tb]
\centering
\includegraphics[width=0.28\textwidth]{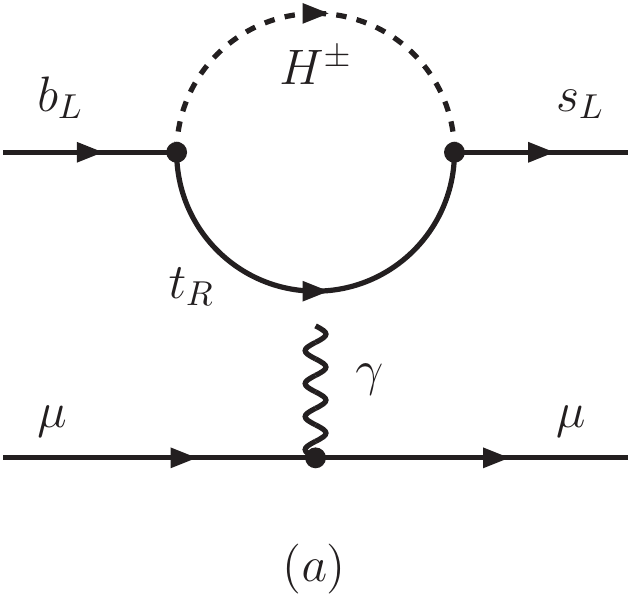} ~~~~~~~~ 
\includegraphics[width=0.28\textwidth]{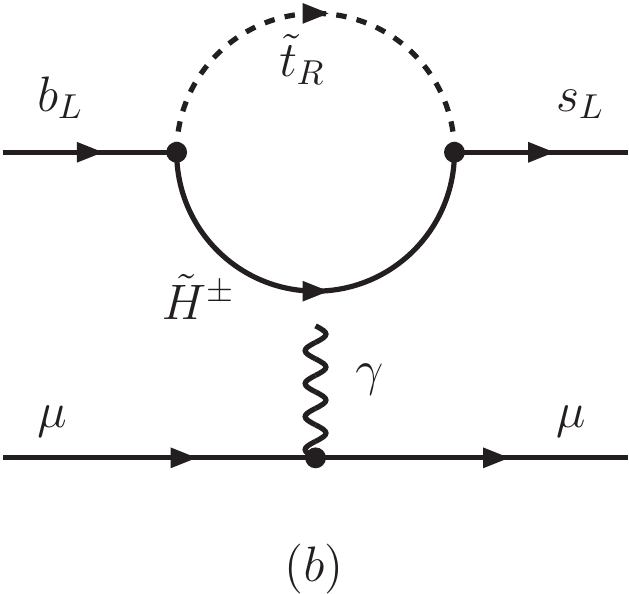} ~~~~~~~~ 
\includegraphics[width=0.28\textwidth]{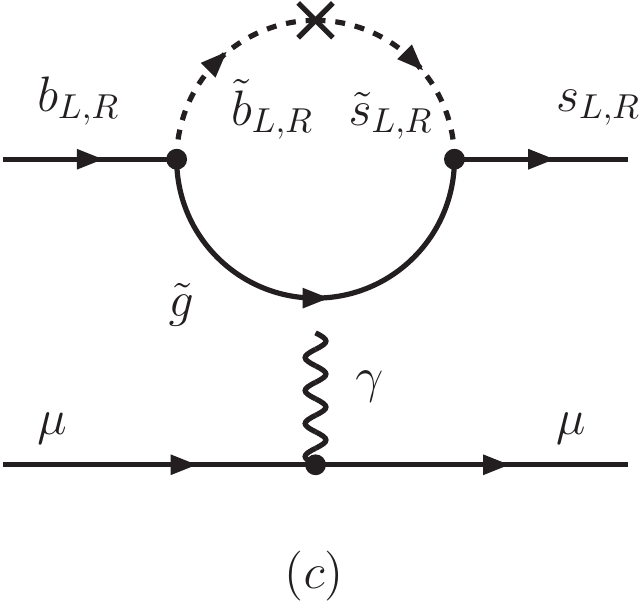} \\[16pt]
\includegraphics[width=0.28\textwidth]{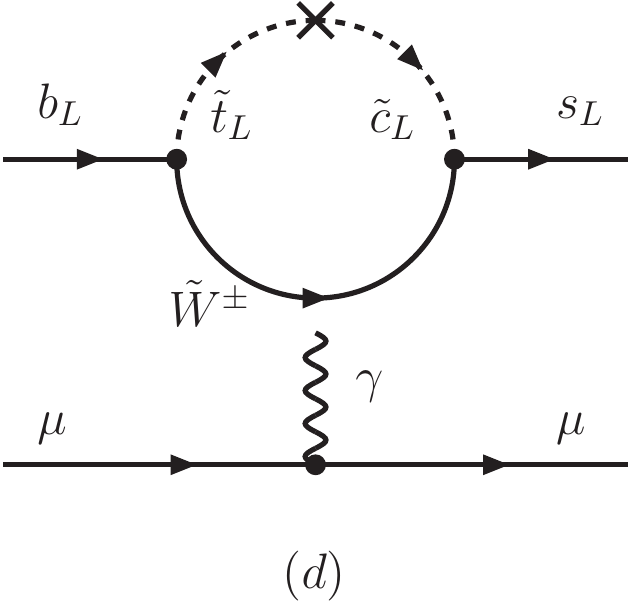} ~~~~~~~~ 
\includegraphics[width=0.28\textwidth]{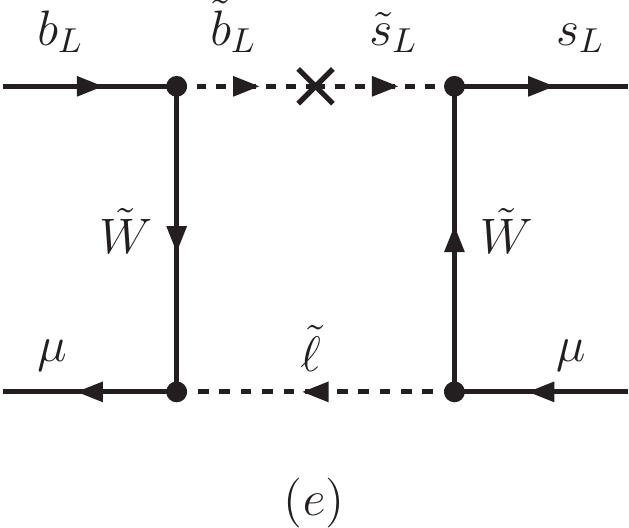}
\caption{Example Feynman diagrams for MSSM contributions to $C_9$ and $C_9^\prime$. Charged Higgs photon penguin (a), Higgsino photon penguin (b), gluino photon penguin (c), charged Wino photon penguin (d), and Wino box (e). In the case of the photon penguins, the photon has to be attached to all charged particles in the loop.}
\label{fig:diagrams}
\end{figure}
\subsubsection{SUSY contributions to \texorpdfstring{$C_{9}^{(\prime)}$}{C9(')}}

Contributions to $C_9$ and $C_9^\prime$ can come from (i) $Z$ penguins, (ii) photon penguins, and (iii) box diagrams.
Example Feynman diagrams for the most important contributions are shown in figure~\ref{fig:diagrams}. 

\paragraph{(i) $Z$ penguin contributions} to the axial-vector coefficients $C_{10}$ and $C_{10}^\prime$ can be sizable in corners of parameter space with large flavour-changing trilinear couplings~\cite{Lunghi:1999uk,Altmannshofer:2009ma}. $Z$ penguin contributions to $C_9$ and $C_9^\prime$, however, are suppressed by the accidentally small vector coupling of the $Z$ to charged leptons $\propto 1 - 4 s_W^2 \simeq 0.08$ and therefore negligible.

\paragraph{(ii) Photon penguin contributions} to $C_9$ and $C_9^\prime$ are induced by charged Higgs, Higgsino, and gaugino loops.
For charged Higgs loops shown in diagram (a) of figure~\ref{fig:diagrams} we find
\begin{equation}
 C_9^{H^\pm} \simeq - \frac{m_t^2}{m_{H^\pm}^2} \frac{1}{24} ~(\cot\beta)^2 ~f_9^{H^\pm}\left(\frac{m_t^2}{m_{H^\pm}^2}\right) \,.
\end{equation}
The loop function can be found in appendix~\ref{loopfunctions} and is normalized such that $f_9^{H^\pm}(1) = 1$.
The charged Higgs loops are strongly suppressed by $(\cot\beta)^2$. Even for extremely small $\tan\beta \sim 1$, we find that this contribution does not exceed $|C_9^{H^\pm}| \lesssim 0.05$ for any value of the charge Higgs mass $m_{H^\pm} > 100$~GeV.
Charged Higgs contributions to the right-handed coefficient $C_9^\prime$ are suppressed by the strange quark mass and also negligible.

Contributions from Higgsino loops are shown in diagram (b) of figure~\ref{fig:diagrams} and lead to 
\begin{equation} \label{eq:C9higgsino}
 C_9^{\tilde H^\pm} \simeq \frac{7}{72} \frac{m_t^2}{m_{\tilde t_R}^2} ~f_9^{\tilde H^\pm}\left(\frac{|\mu|^2}{m_{\tilde t_R}^2}\right)  \,.
\end{equation}
The loop function $f_9^{\tilde H^\pm}$ is reported in the appendix~\ref{loopfunctions} and normalized such that for a Higgsino that is degenerate with the right-handed stop, $|\mu| \simeq m_{\tilde t_R}$, one has $f_9^{\tilde H^\pm}(1) = +1$. The loop function remains positive for a wide range of Higgsino and stop masses. A negative sign for $C_9^{\tilde H^\pm}$, as required to address the discrepancies in $B \to K^* \mu^+\mu^-$, can only be obtained for heavy Higgsino masses $|\mu| \gtrsim 5 m_{\tilde t_R}$. This region of parameter space implies $f_9^{\tilde H^\pm} \ll 1$ and $C_9^{\tilde H^\pm}$ is completely negligible.

Evaluating gluino loops (diagram (c) of figure~\ref{fig:diagrams}), we find contributions to both $C_9$ and $C_9^\prime$
\begin{equation}
 (V_{ts}^* V_{tb})~\Big\{ C_9^{\tilde g} , C_9^{\prime\,\tilde g} \Big\} \simeq -\frac{8}{135} \frac{g_s^2}{g_2^2} \frac{m_W^2}{m_{\tilde d}^2} ~\Big\{ (\delta_{bs}^L) , (\delta_{bs}^R) \Big\}~f_9^{\tilde g}\left(\frac{m_{\tilde g}^2}{m_{\tilde d}^2}\right) \,,
\end{equation}
where for simplicity we assume a common mass $m_{\tilde d}$ for all down-type squarks and the loop function, that can be found in the appendix~\ref{loopfunctions}, is normalized such that $f_9^{\tilde g}(1) = 1$. The phase of $C_9^{\tilde g}$ and $C_9^{\prime\,\tilde g}$ is set by the phase of the squark mixing $\delta_{bs}^L$ and $\delta_{bs}^R$, that are free parameters.
Despite the enhancement by the strong coupling constant and the possible enhancement by $O(1)$ squark mixing angles, we find that this contribution is negligible. Sizable contributions $|C_9^{\tilde g}| \gtrsim 0.5$ are only possible for extremely light squarks and gluinos of order 200 GeV that are completely excluded by direct searches. Similar conclusions can be drawn for neutral Wino and Bino loops, that, compared to the gluino loops, are further suppressed by small electro-weak gauge couplings.

Finally there are charged Wino loops that can induce photon penguins. We find
\begin{equation}
 (V_{ts}^* V_{tb})~C_9^{\tilde W} \simeq - \frac{1}{45} \frac{m_W^2}{m_{\tilde d}^2} ~(\delta_{bs}^L)~f_9^{\tilde W}\left(\frac{m_{\tilde W}^2}{m_{\tilde d}^2}\right) \,,
\end{equation}
where the loop function given in the appendix~\ref{loopfunctions}.
For a degenerate spectrum, $f_9^{\tilde W}(1) = 1$, and the contribution is again negligible. For a large hierarchy between the Wino and down-type squark masses $m_{\tilde W} \ll m_{\tilde d}$, however, the contribution is enhanced by a large logarithm $f_9^{\tilde W}(x) \xrightarrow{x \to 0} - 30 \log(m_{\tilde W}^2/m_{\tilde d}^2)$. The large logarithm arises from the diagram where the photon attaches to the light charged Wino in the loop (see diagram (d) of figure~\ref{fig:diagrams}) and is therefore absent in the neutral gaugino loops discussed above.
Despite the large logarithm, we find that appreciable contributions of the order of $|C_9^{\tilde W}| \gtrsim 0.5$ are only possible for maximal squark mixing $\delta_{bs}^L \sim O(1)$, Winos close to the LEP bound $m_{\tilde W} \sim 100$~GeV and bottom and strange squarks with a mass of $m_{\tilde d} \sim 400$~GeV. Such light squarks are not only in tension with strong bounds from direct searches for sbottoms~\cite{Chatrchyan:2013lya,ATLAS-CONF-2013-053}, but in combination with the $O(1)$ squark mixing would generically also lead to too large contributions to other well measured flavour observables, like $B_s$ mixing.

\paragraph{(iii) Box contributions} to $C_9$ and $C_9^\prime$ are again induced by charged Higgs, Higgsino, and gaugino loops.
Charged Higgs boxes and Higgsino boxes are suppressed by the tiny muon Yukawa coupling and thus completely negligible.
For the contribution from Wino boxes (diagram (e) of figure~\ref{fig:diagrams}) we find 
\begin{equation}
 (V_{ts}^* V_{tb})~C_9^\text{box} \simeq \frac{1}{s_W^2} \frac{5}{192} \frac{m_W^2}{m_{\tilde d}^2} ~(\delta_{bs}^L)~ f_9^\text{box}\left(\frac{m_{\tilde \ell}^2}{m_{\tilde d}^2},\frac{m_{\tilde W}^2}{m_{\tilde d}^2}\right) \,.
\end{equation}
The loop function $f_9^\text{box}$ is given in the appendix~\ref{loopfunctions}.
In the limit of degenerate Wino, slepton and down-type squark masses $m_{\tilde W} = m_{\tilde \ell} = m_{\tilde d}$, the loop function reduces to $f_9^\text{box}(1,1) = 1$. For an approximate degenerate SUSY spectrum, we find again tiny contributions $C_9^\text{box} \ll 0.1$ even for down-squarks as light as 500~GeV and assuming maximal squark mixing $\delta^L_{bs} \sim O(1)$.
The same conclusion holds for the Bino boxes and mixed Wino-Bino boxes that are further suppressed by the small hypercharge gauge coupling.
For a hierarchical spectrum, i.e. if both Winos and sleptons are light compared to the squarks, $m_{\tilde W}, m_{\tilde \ell} \ll m_{\tilde d}$, the box contributions get enhanced by large logarithms $\log(m_{\tilde W}/m_{\tilde d})$, $\log(m_{\tilde \ell}/m_{\tilde d})$. Assuming for simplicity $m_{\tilde \ell} = m_{\tilde W}$, we find $f_9^\text{box}(x,y) \xrightarrow{x,y \to 0} -12 \log(m_{\tilde \ell}^2/m_{\tilde d}^2)$. Similar to the charge Wino photon penguins discussed above, 
maximal squark mixing, Winos and leptons close to the LEP bound $\sim 100$~GeV as well as light bottom and strange squarks with a mass of $m_{\tilde d} \sim 500$~GeV are required to reach $|C_9^\text{box}| \gtrsim 0.5$. Such a region of parameter space is strongly disfavored by direct searches~\cite{Chatrchyan:2013lya,ATLAS-CONF-2013-053}, and other flavour constraints.
Analogous comments apply to Bino and mixed Wino-Bino boxes with a hierarchical spectrum.

\subsubsection{SUSY contributions to \texorpdfstring{$C_{7}^{(\prime)}$}{(')}}

Sizable contributions to the dipole operators, on the other hand, can be generated in various ways in the MSSM. We discuss them separately in the following, keeping in mind that they could also be present simultaneously. 
If the charged Higgs boson is light, it can contribute significantly to the Wilson coefficient $C_7$. At one loop one finds
\begin{equation} \label{eq:C7higgs}
 C_7^{H^\pm} \simeq - \frac{7}{36} \frac{m_t^2}{m_{H^\pm}^2} ~f_7^{H^\pm}\left(\frac{m_t^2}{m_{H^\pm}^2}\right) \,,
\end{equation}
with the loop function given in the appendix~\ref{loopfunctions}.
In (\ref{eq:C7higgs}) we neglected a term that is proportional to $(\cot\beta)^2$ and only relevant for very small $\tan\beta$.
The function $f_7^{H^\pm}$ is always positive, and therefore the charged Higgs loop always interferes constructively with the SM contribution $C_7^{\rm SM} \sim -0.31$.
A charged Higgs contribution of $C_7^{H^\pm} \simeq -0.07$, as in scenario (7) of table~\ref{tab:scen}, corresponds to a charged Higgs mass of $m_{H^\pm}^2 \simeq 500$~GeV.

Large contributions to $C_7$ can also arise from Higgsino--stop loops. The leading term reads
\begin{equation}
 C_7^{\tilde H^\pm} \simeq \frac{5}{72} \frac{m_t^2}{m_{\tilde t_R}^2} \frac{\mu A_t}{m_{\tilde t_R}^2} ~\tan\beta ~f_7^{\tilde H^\pm}\left(\frac{|\mu|^2}{m_{\tilde t_R}^2}\right) \,.
\end{equation}
The sign of $C_7^{\tilde H^\pm}$ is determined by sign$(\mu A_t)$, and the loop function, that can be found in the appendix, is normalized such that $f_7^{\tilde H^\pm}(1)=1$. Note that GUT-scale boundary conditions typically lead to constructive interference of the chargino and SM contributions if $\mu$ is negative.
For illustration we now work in the limit where the right-handed stop and the Higgsino are approximately degenerate $m_{\tilde t_R} \simeq |\mu|$. Assuming also a sizable trilinear coupling $A_t \simeq m_{\tilde t_R}$, the right magnitude of $C_7^{\tilde H^\pm}$ as in scenario (7) of table~\ref{tab:scen}, can be achieved with light stops with mass $m_{\tilde t_R} \sim 500$~GeV and a moderate $\tan\beta \sim 10$. For large values of $\tan\beta \sim 50$, stops can be as heavy as $1.2$~TeV. 

Note that the charged Higgs and Higgsino contributions do not require any flavour structure beyond the SM CKM matrix. They only contribute to the left-handed dipole coefficient $C_7$.
If there is non-trivial squark mixing, also gaugino contributions can become relevant and generate both $C_7$ (gluino, Wino, and Bino) and $C_7^\prime$ (gluino and Bino). Due to the large strong coupling constant, gluino contributions are typically dominant. The leading gluino contributions read
\begin{equation}
 (V_{ts}^*V_{tb})~\Big\{ C_7^{\tilde g} , C_7^{\prime\,\tilde g} \Big\} \simeq - \frac{2}{45} \frac{g_s^2}{g_2^2} \frac{m_W^2}{m_{\tilde d}^2} \frac{\mu m_{\tilde g}}{m_{\tilde d}^2} ~\tan\beta~ \Big\{\delta^L_{bs} , \delta^R_{bs} \Big\}~f_7^{\tilde g}\left(\frac{m_{\tilde g}^2}{m_{\tilde d}^2}\right) \,.
\end{equation}
The above expression assumes a common down-type squark mass $m_{\tilde d}^2$. For approximately degenerate gluinos and down-type squarks one has $f_7^{\tilde g}(1) = 1$. In regions of parameter space where the Higgsino mass is sizable, $|\mu| \sim m_{\tilde d}$, squark mixing is of O(1), $|\delta_{bs}^L| \simeq 0.3$, $|\delta_{bs}^R| \simeq 0.5$, and $\tan\beta$ is large, $\tan\beta \sim 50$, down squarks and gluinos at $\sim 2$~TeV lead to interesting values for the Wilson coefficients $C_7^{\tilde g} \simeq -0.06$ and $C_7^{\prime \,\tilde g} \simeq -0.10$ as in scenario (77') of table~\ref{tab:scen}.

\subsubsection{Summary: MSSM}
In summary, we systematically discussed all possible 1-loop contributions to $C_9$ and $C_9^\prime$ in the MSSM and found that even allowing for generic $O(1)$ squark mixing, $C_9$ and $C_9^\prime$ remain to a good approximation SM-like. The tensions in the $B \to K^* \mu^+\mu^-$ data can therefore only be softened slightly by modifications of the dipole Wilson coefficient $C_7$ and $C_7^\prime$, that can arise in a variety of MSSM models with a TeV scale spectrum.

\subsection{Models with partial compositeness} \label{sec:PC}

Next to supersymmetry, the most well-motivated class of models solving the gauge hierarchy problem are models with a composite Higgs boson or extra-dimensional models dual to four-dimensional composite Higgs models.
In all these models, generating fermion masses without generating excessive FCNCs requires, from the 4D perspective, the mechanism of partial compositeness, giving mass to fermions by linear mixing with composite operators. A simple setup to obtain approximate results in this framework is given by the two-site description of ref.~\cite{Contino:2006nn}, describing one composite resonance for each SM field (plus an extended spin-1 sector to accommodate a custodial symmetry required by electroweak precision tests).

Similarly to the MSSM, effects in dipole operators are easily generated in these models, while effects in $C_9$ or $C_9'$ are difficult to generate. We will discuss both cases in turn.

\subsubsection{Partial compositeness and \texorpdfstring{$C_9^{(\prime)}$}{C9(')}}

FCNCs can arise already at the tree level and are mediated either by the $Z$ boson (which is partially composite but mostly elementary) or by heavy vector resonances that we collectively denote by $\rho$.

\begin{figure}[tp]
\includegraphics[viewport=30 620 360 820,clip,width=0.31\textwidth]{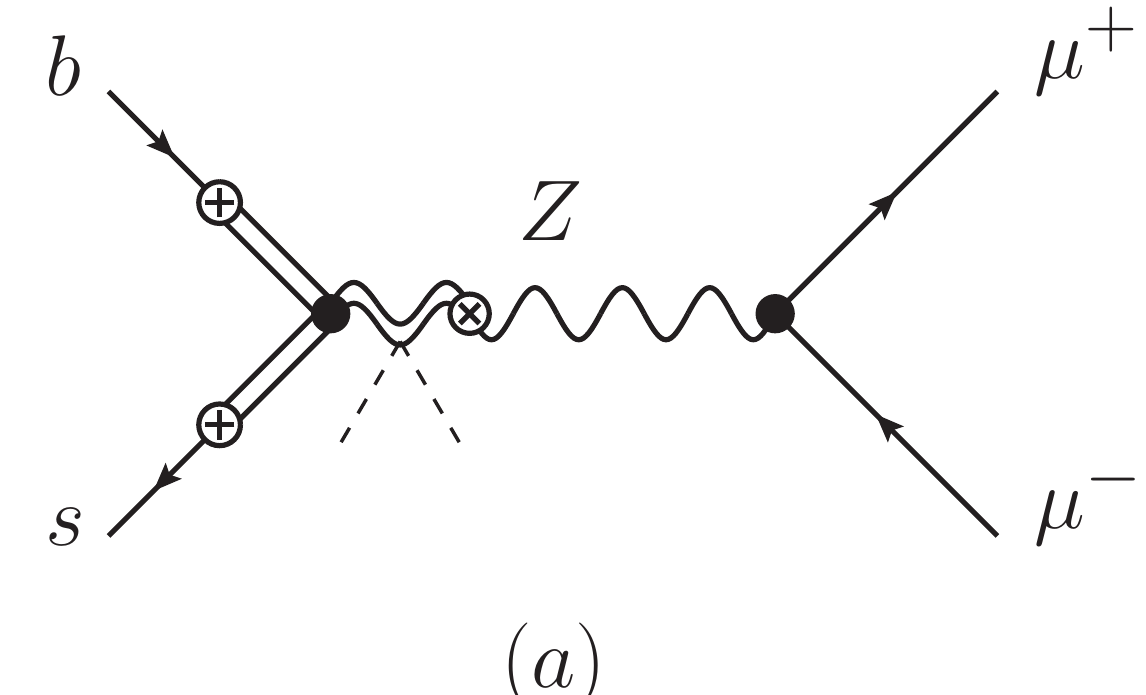}%
\hspace{0.03\textwidth}%
\includegraphics[viewport=30 620 360 820,clip,width=0.31\textwidth]{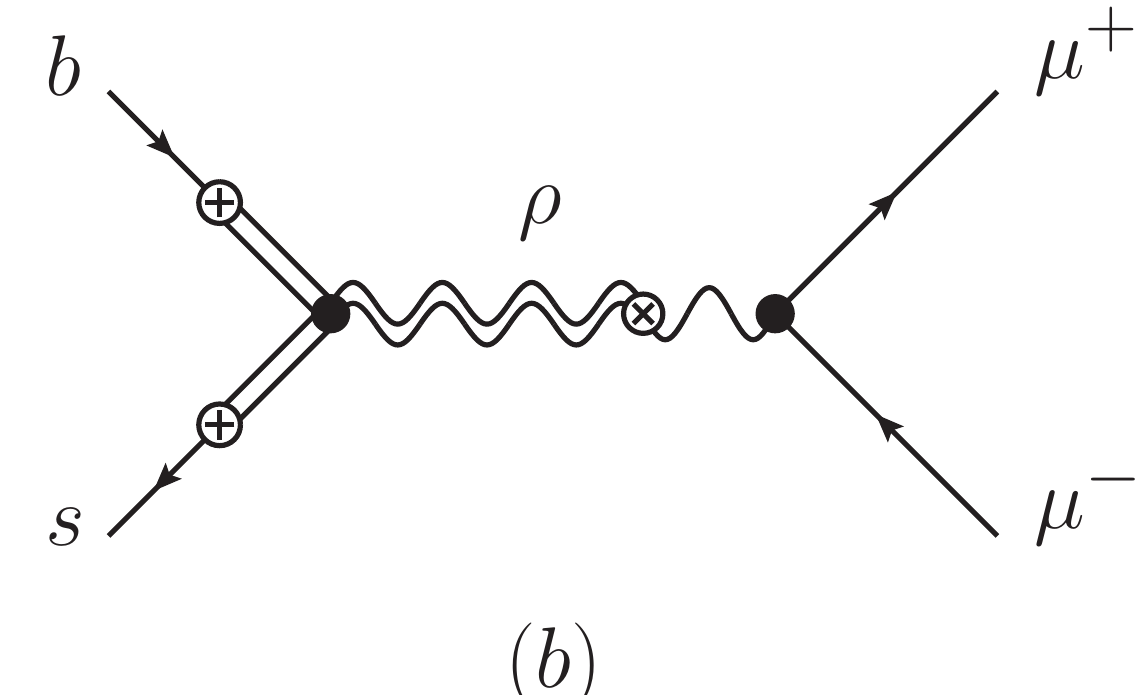}%
\hspace{0.03\textwidth}%
\includegraphics[viewport=30 620 360 820,clip,width=0.31\textwidth]{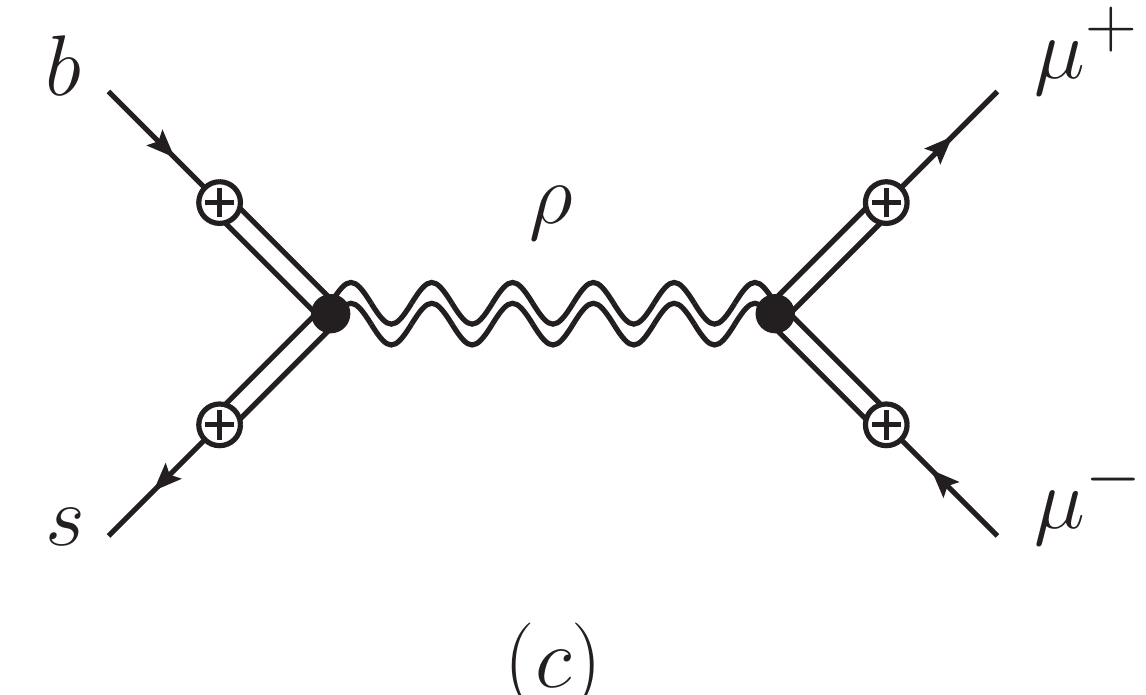}%
\caption{Feynman diagrams for tree-level contributions to $C_9$ and $C_9^\prime$ in models with partial compositeness. Double lines stand for composite fields; the dashed lines in diagram (a) indicate that two Higgs VEV insertions are required.}
\label{fig:diagrams_PC}
\end{figure}

\paragraph{(i) $Z$ exchange}
As is well known, tree-level flavour-changing $Z$ couplings can lead to sizable contributions to $C_{10}$ and $C_{10}'$ in these models \cite{Agashe:2004ay,Burdman:2002gr,Blanke:2008yr,Bauer:2009cf}.
They arise after electroweak symmetry breaking from the mixing of fermions or vector bosons with different electroweak quantum numbers and can be estimated as
\begin{align}
\frac{e^2}{16\pi^2} (V_{ts}^*V_{tb}) \, C_{10}^{(\prime)Z}
&\sim
\left(
a_{L,R}\,\frac{v^2Y^2}{2m_\psi^2}
+
b_{L,R}\,\frac{v^2g_\rho^2}{4m_\rho^2}
\right)
s_{L,R}^bs_{L,R}^s
\,,
\end{align}
where $s_{L,R}$ are the degrees of compositeness of left- and right-handed quarks, $m_{\psi,\rho}$ are composite fermion and vector resonance masses, $Y$ is a strong Yukawa coupling and $a_{L,R},b_{L,R}$ are $O(1)$ numbers that depend on the fermion representations. In presence of a custodial protection of $Z$ couplings, either the $a_L,b_L$ or  the $a_R,b_R$ vanish \cite{Agashe:2006at,Blanke:2008yr,Straub:2013zca}. The flavour-changing $Z$ couplings also contribute to $C_9$ and $C_9'$, but as in the SUSY case, they are suppressed by the accidentally small vector coupling of the $Z$ to charged leptons, such that
\begin{equation}
C_9^{(\prime)Z} \approx -0.08\,C_{10}^{(\prime)Z}
\,.
\end{equation}
A recent numerical analysis of different models \cite{Straub:2013zca}, taking into account tree-level $\Delta F=2$ and electroweak constraints, found maximal contributions of about $|C_{10}^{(\prime)Z}|\lesssim2.5$, which implies $|C_{9}^{(\prime)Z}|\lesssim0.2$, too small to solve the tensions in $B\to K^*\mu^+\mu^-$.

\paragraph{(ii) $\rho$ exchange}
Another source of tree-level contributions to the semi-leptonic operators comes from the exchange of neutral heavy vector resonances. This is a special case of the flavour-changing $Z'$ discussed above, so we can use the same notation. For the flavour-changing coupling, one expects
\begin{align}
\frac{g_2}{2c_W} g^{L,R}_{bs} &\sim
g_\rho \, s_{L,R}^bs_{L,R}^s 
\end{align}
and we stress that it is non-zero even in the limit of vanishing Higgs VEV, in contrast to the $Z$-mediated contribution.
For the coupling to muons, one has to distinguish between two contributions. In the first case, diagram (b) of figure~\ref{fig:diagrams_PC}, the coupling proceeds through an admixture to $\rho$ of an elementary $Z$ boson. In that case, the vector and axial vector couplings of the $\rho$ are equal to the vector and axial vector couplings of the SM $Z$ to muons. Consequently, also here the contributions to $C_9^{(\prime)}$ are strongly suppressed. In the second case, shown in diagram (c) of figure~\ref{fig:diagrams_PC}, the $\rho$ directly couples to the composite component of the muons. In this case, the couplings depend on the degrees of compositeness of left- and right-handed muons and  can be different from the $Z$ couplings,
\begin{align}
g^{V}_{\mu} &\sim
g_\rho \left[(s_{L}^\mu)^2+(s_{R}^\mu)^2\right]
\,,
&
g^{A}_{\mu} &\sim
g_\rho \left[(s_{L}^\mu)^2-(s_{R}^\mu)^2\right]
\,.
\end{align}
Due to the lightness of the muon, it is expected to be mostly elementary and the above couplings should be strongly suppressed. They could only be sizable in the extreme case where one chirality of muons has a large degree of compositeness. Then however, one has $g^{V}_{\mu}\simeq \pm g^{A}_{\mu}$, since the product $s_{L}^\mu s_{R}^\mu$ has to be tiny.
In other words, it is not possible to obtain a sizable contribution to $C_9$ without at the same time generating a large correction to $C_{10}$, which is disfavoured by the data.

To quantify the last statement, we can make a fit to the data as in section~\ref{sec:fit}, but fixing $C_9^\text{NP}=\pm C_{10}^\text{NP}$.
In the case of equal sign, we find $\Delta \chi^2=0.8$, so no significant improvement. In the case of opposite sign (corresponding to a composite $\mu_R$), we find $\Delta \chi^2=2.3$ for $C_9^\text{NP}=-0.3$, corresponding to a very small improvement. For $C_9'=\pm C_{10}'$, the improvement is even smaller.

\subsubsection{Partial compositeness and \texorpdfstring{$C_7^{(\prime)}$}{C7(')}}

Dipole operators are generated at the one loop level in partial compositeness \cite{Agashe:2008uz,Gedalia:2009ws,Blanke:2012tv,Vignaroli:2012si}. The dominant contribution typically comes from a diagram with a Higgs and a heavy, vector-like fermion in the loop, since the large fermion mass lifts the chirality suppression of the amplitude.
One can write it generically as
\begin{align}
(V_{ts}^*V_{tb}) \,m_b\, C_{7}^{(\prime)}
\sim
s_{R,L}^b s_{L,R}^s
\frac{v^3\, Y\tilde Y Y}{\sqrt{2} \, m_\psi^2}
\frac{1}{12}
f_7^h\left( \frac{m_\psi^2}{m_h^2} \right),
\end{align}
where we suppressed the flavour structure and a model-dependent $O(1)$ overall factor.
Here, $\tilde Y$ is a ``wrong-chirality'' Yukawa coupling in the strong sector.
A similar contribution also comes from
loops with a $W$ or $Z$ boson instead of the Higgs.
In the flavour-anarchic model, one can estimate
\begin{align}
s_{R}^b s_{L}^s
\frac{v^3\, Y\tilde Y Y}{\sqrt{2} \, m_\psi^2}
&\sim
(V_{ts}^*V_{tb}) \,m_b\, \frac{v^2Y\tilde Y}{m_\psi^2}
\,,
&
s_{L}^b s_{R}^s
\frac{v^3\, Y\tilde Y Y}{\sqrt{2} \, m_\psi^2}
&\sim
\frac{m_s}{(V_{ts}^*V_{tb})} \frac{v^2Y\tilde Y}{m_\psi^2}
\,.
\end{align}
Since  $m_s/(m_b V_{ts}^2)\approx15$, one generically expects larger contributions to $C_7'$ than to $C_7$. Taking for example $Y\sim \tilde Y \sim 3$ and $m_\psi\sim 1$~TeV, one finds 
$|C_7|\sim 0.05, |C_7'|\sim0.7$. We stress however that these estimates are subject to sizable corrections since we neglected various $O(1)$ factors throughout.

\subsubsection{Summary: partial compositeness}

To summarize, in models with partial compositeness one generically expects NP contributions to $C_7$ and $C_7'$ that are of the right size to reproduce scenario $(77')$ above and thus ameliorate the tensions in $B\to K^*\mu^+\mu^-$. Generating a sizable contribution to $C_9$ or $C_9'$, which is required to fully remove the tensions, requires a large degree of compositeness for one chirality of muons as well as a cancellation between several contributions to $C_{10}$ and/or $C_{10}'$. Whether such scenario is viable when taking into account constraints on the lepton sector is an interesting question for future study.

\subsection{Expectations for CP Asymmetries}\label{sec:CP}

\begin{figure}[tp]
\centering
\includegraphics[width=0.49\textwidth]{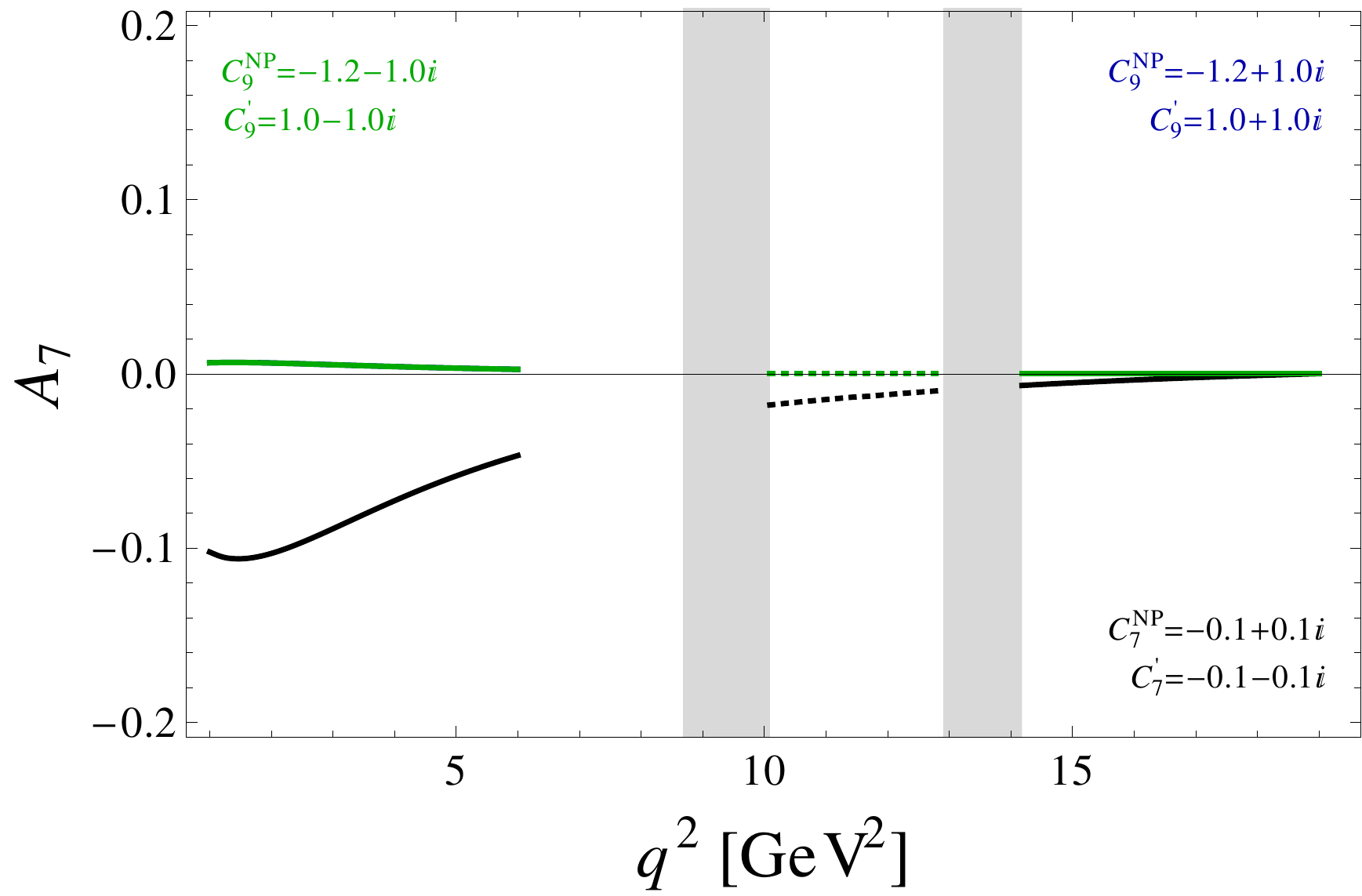} \includegraphics[width=0.49\textwidth]{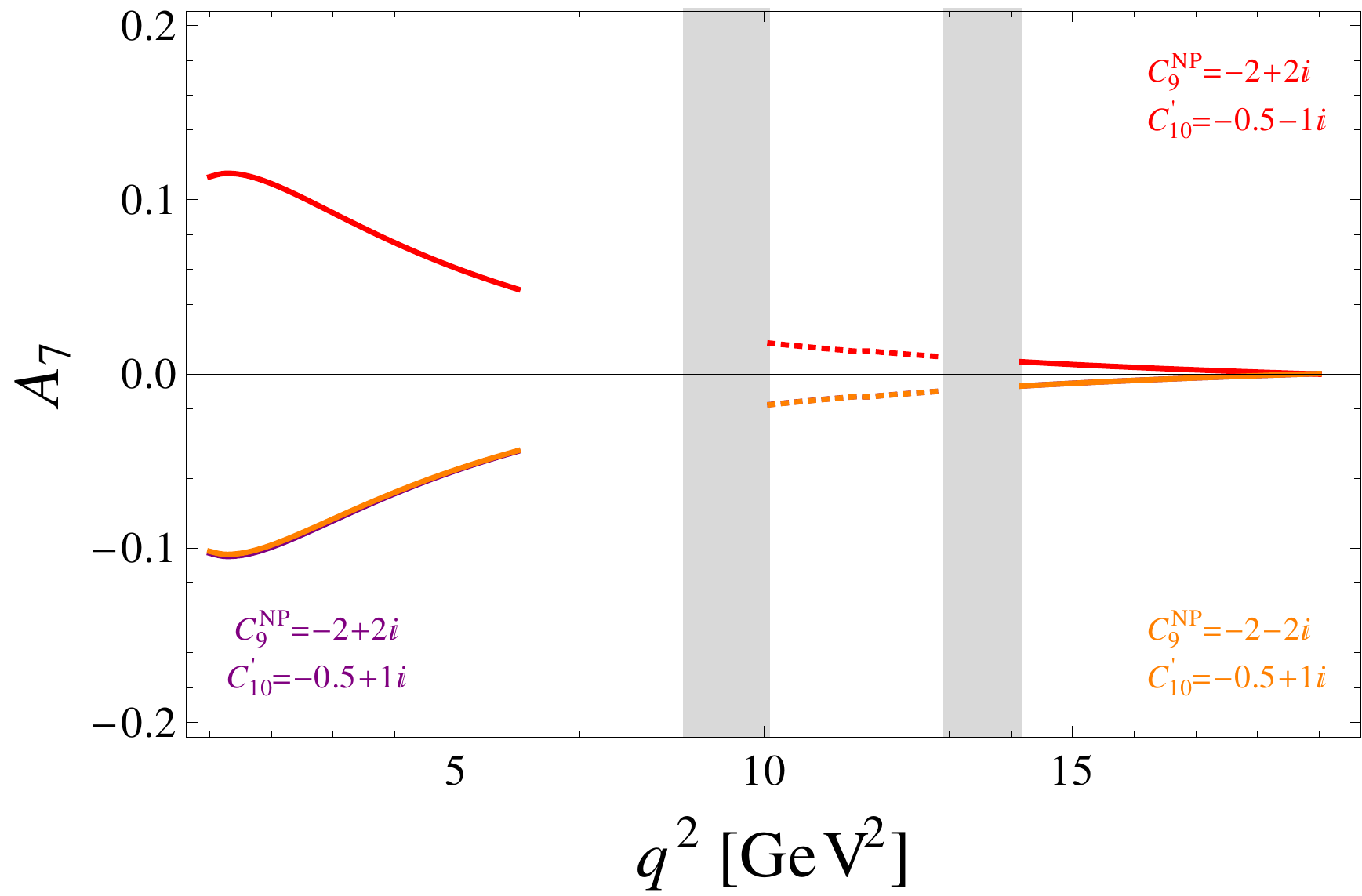} \\[16pt]
\includegraphics[width=0.49\textwidth]{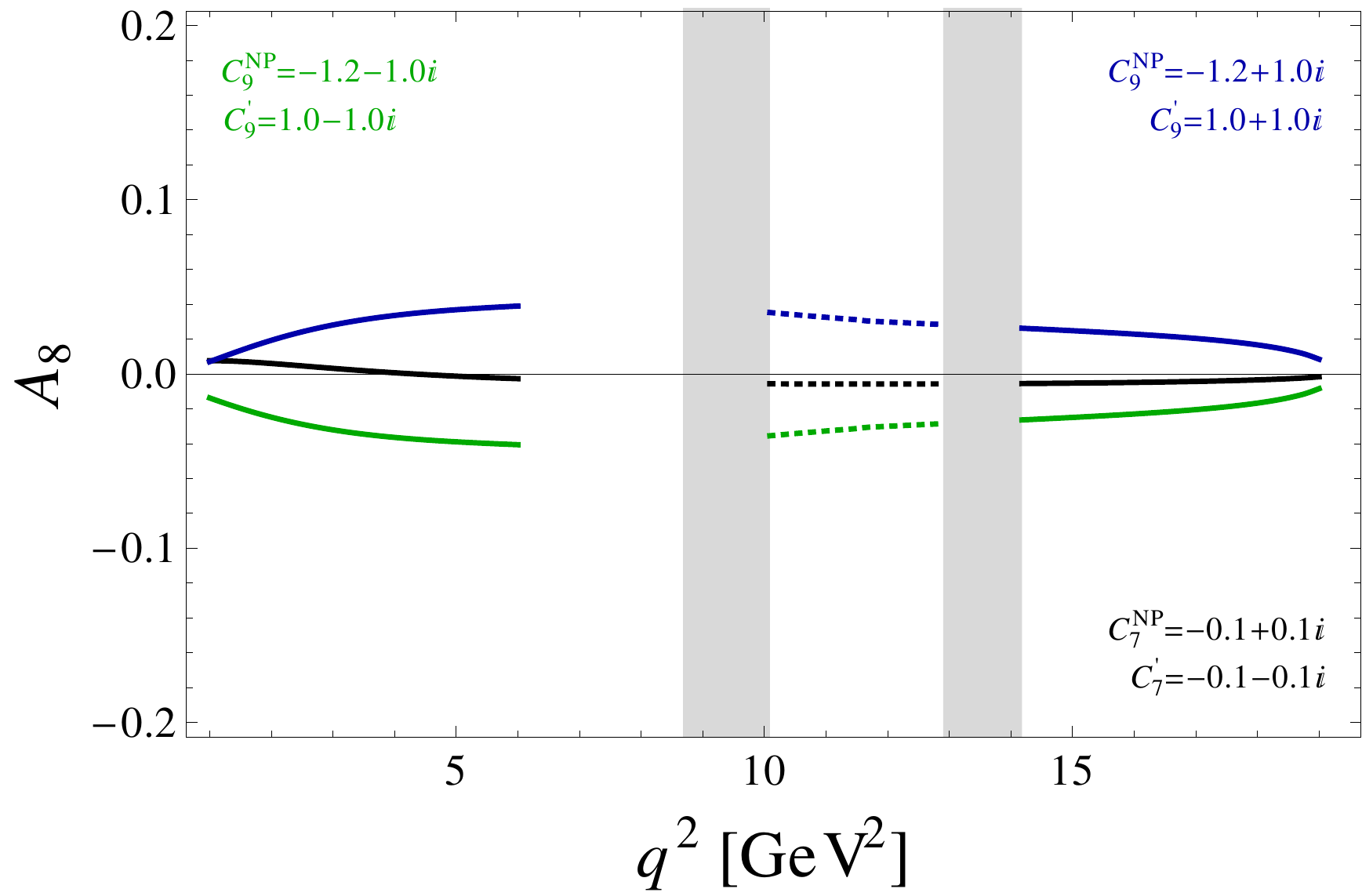} \includegraphics[width=0.49\textwidth]{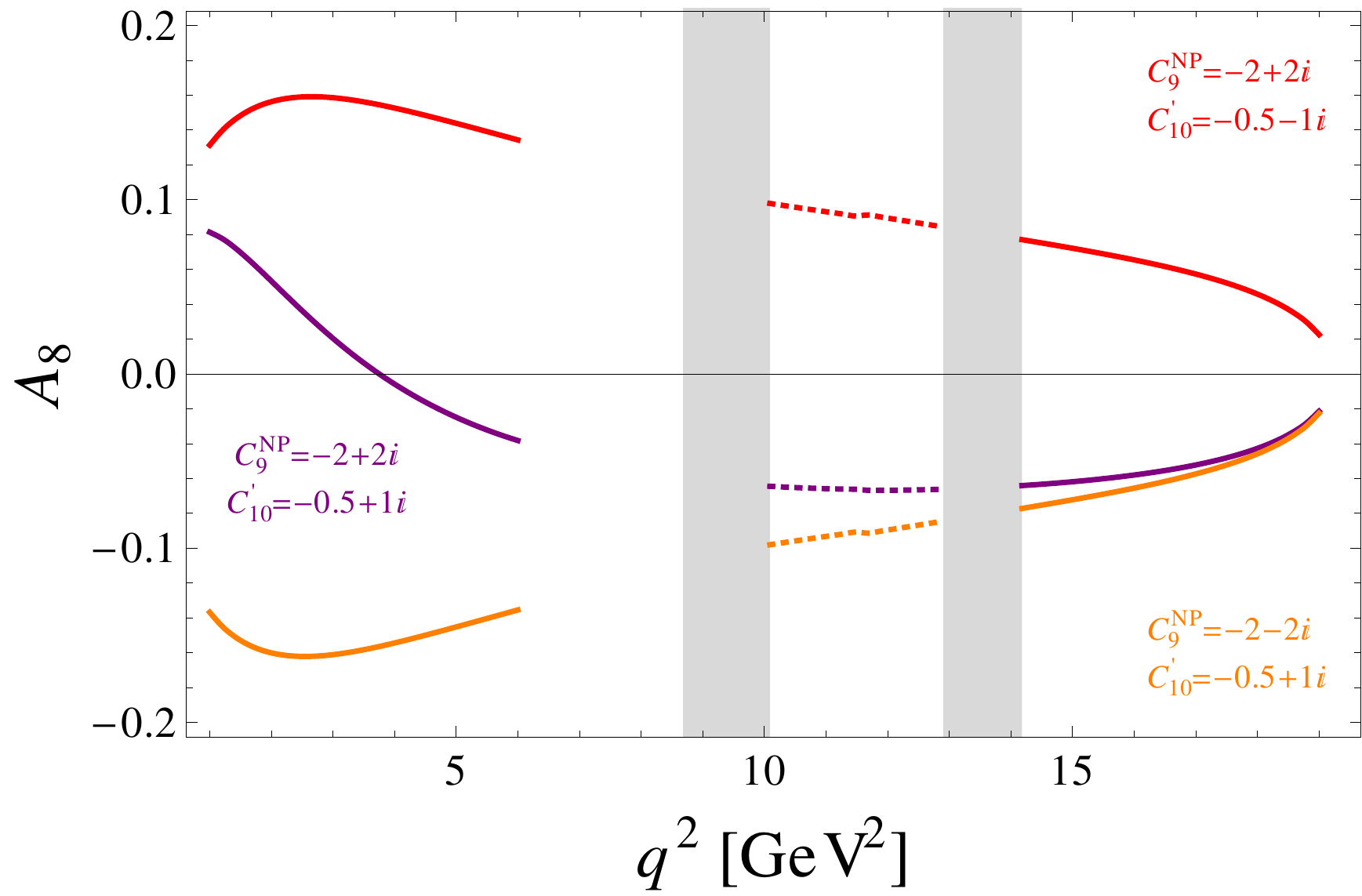} \\[16pt]
\includegraphics[width=0.49\textwidth]{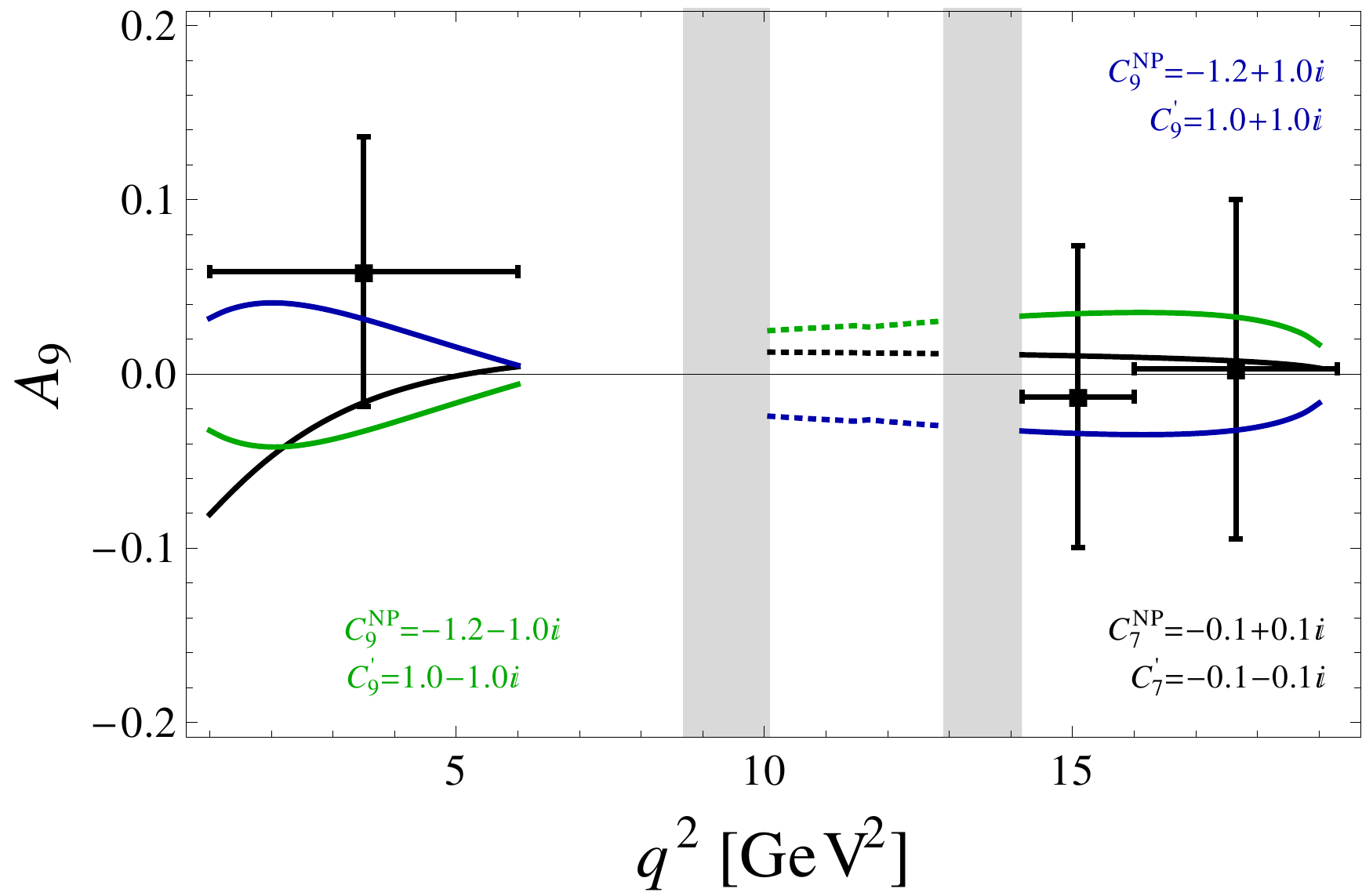} \includegraphics[width=0.49\textwidth]{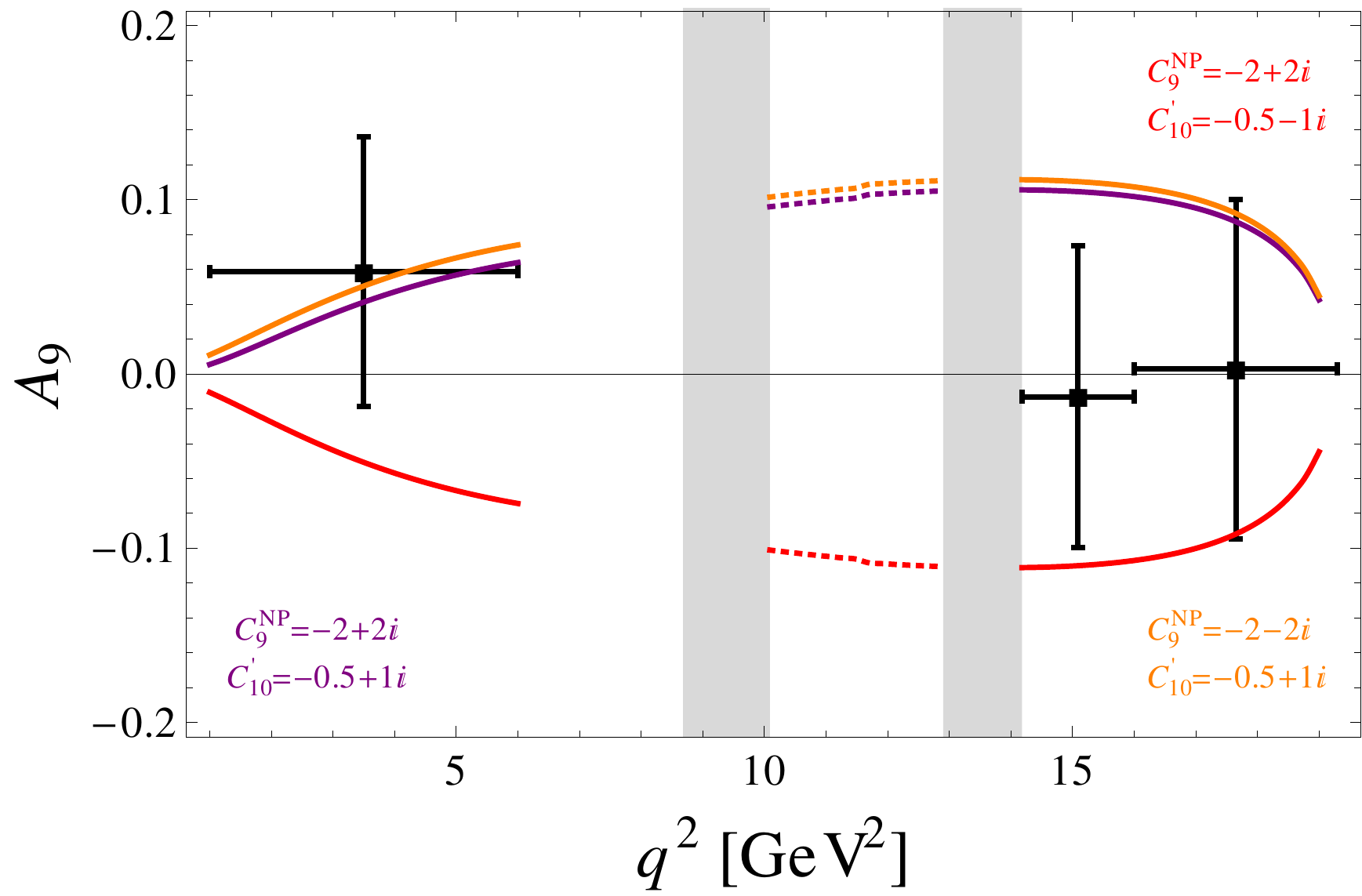}
\caption{Predictions for the CP asymmetries $A_7$, $A_8$ and $A_9$ as function of the di-muon invariant mass squared $q^2$ in various scenarios that address the observed discrepancies in $B \to K^* \mu^+\mu^-$. The values for the Wilson coefficients corresponding to each scenario are indicated explicitly in the plots. SM predictions for the CP asymmetries are negligibly small throughout the whole $q^2$ range.}
\label{fig:ACP}
\end{figure}

Although all the tensions in the data occur in CP-averaged observables and we therefore stressed above that NP effects required to remove them may be aligned in phase with the SM, this is mostly due to the fact that few CP asymmetries have been measured to a good precision and the imaginary parts of the Wilson coefficients are still poorly constrained (cf.~\cite{Altmannshofer:2012az}).
Generically however, without imposing additional restrictions on new sources of CP violation, most of the discussed NP contributions to the Wilson coefficients are expected to be complex. Under the generic assumption that the imaginary parts are of the same order as the real parts, we can derive generic expectations for the CP asymmetries $A_7$, $A_8$ and $A_9$, in the considered scenarios that address the tensions in the data.\footnote{We explicitly checked that in all cases discussed below and shown in fig~\ref{fig:ACP}, the presence of imaginary parts does not worsen the agreement with the data significantly.}

We provide simple approximate expressions for the T-odd CP asymmetries at low $q^2$
\begin{align}
\langle A_7 \rangle_{[1,6]}
\simeq &
-0.44 \,\text{Im}(C_{7}^\text{NP})
+0.44 \,\text{Im}(C_{7}^\prime)
+0.07\,\text{Im}(C_{10}^\text{NP})
-0.07\,\text{Im}(C_{10}^\prime) ~,
\nonumber\\
\langle A_8 \rangle_{[1,6]}
\simeq &
+0.25 \,\text{Im}(C_{7}^\text{NP})
+0.23 \,\text{Im}(C_{7}^\prime)
+0.04\,\text{Im}(C_{9}^\text{NP})
+0.02\,\text{Im}(C_{9}^\prime)
-0.06\,\text{Im}(C_{10}^\prime) ~,
\nonumber\\
\langle A_9 \rangle_{[1,6]}
\simeq &
+0.12 \,\text{Im}(C_{7}^\prime)
+0.04\,\text{Im}(C_{10}^\prime) ~.
\end{align}

The plots in figure~\ref{fig:ACP} show the $q^2$ distributions of the CP asymmetries $A_7$, $A_8$ and $A_9$ in various NP scenarios.
The black curves in the plots on the left hand side correspond to the scenario with non-zero $C_7$ and $C_7^\prime$ of figure~\ref{fig:BKsll}, modified to include imaginary parts of the Wilson coefficients that are not excluded by the data on $S_{K^*\gamma}$ (the time dependent CP asymmetry in $B \to K^* \gamma$) and $A_\text{CP}(b \to s\gamma)$ (the direct CP asymmetry in $B \to X_s \gamma$). Measurements of $A_7$ and $A_9$ at low $q^2$ are sensitive to such a scenario.
The green and blue curves in the plots on the left hand side are similar to the scenario with non-zero $C_9$ and $C_9^\prime$ of figure~\ref{fig:BKsll}. Here, imaginary parts of $O(1)$ with different signs are switched on. Moderate effects in $A_8$ and $A_9$ at the level of $\sim5\%$ are expected in this case.
Finally, the purple, red and orange curves in the plots on the right hand side correspond to a scenario where the tensions in $B \to K^* \mu^+\mu^-$ are explained by NP in $C_9$ and $C_{10}^\prime$. The shown choices of the imaginary parts lead to sizable effects in all three CP asymmetries of the order of $10\%$--$15\%$.

We stress again
that the currently observed tensions in $B \to K^* \mu^+\mu^-$ 
are all confined to CP-averaged observables that
are hardly sensitive to CP phases. It is therefore not possible to predict the sign or the exact size of the expected CP asymmetries
Nevertheless, the generic examples shown in figure~\ref{fig:ACP} demonstrate that precise measurements of the CP asymmetries would allow to further narrow down possible NP explanations.

\section{Conclusions}\label{sec:concl}

Confronting predictions for $B \to K^* \mu^+\mu^-$ angular observables with recent measurements by ATLAS, CMS and LHCb, we pointed out three discrepancies at the 2--3$\sigma$ level, namely in the observables $F_L$ and $S_5$ at low $q^2$ and in the observable $S_4$ at high $q^2$. We performed a model-independent analysis taking into account all relevant constraints, finding in particular that
\begin{enumerate}
\item the tension in $S_4$ cannot be resolved by NP contributions and is therefore likely due to statistical fluctuations or underestimated errors (with the resonance contributions mentioned in section~\ref{sec:fit} a possible source for the latter);
\item the tension in $S_5$ and $F_L$ can be reduced by a negative NP contribution to the Wilson coefficient $C_9$ only, but measurements of BR($B\to K\mu^+\mu^-$) and $A_\text{FB}$ prevent a complete solution;
\item allowing simultaneous NP contributions to two Wilson coefficients, there are various possibilities summarized in table~\ref{tab:scen}. The best fit is obtained with simultaneous NP contributions to $C_9$ and $C_9'$ with opposite sign. Modifying the dipole operator coefficients $C_7$ and $C_7'$ only, the tensions can be reduced.
\end{enumerate}

Very recently, a model independent analysis of $B \to K^* \mu^+\mu^-$ appeared \cite{Descotes-Genon:2013wba}, that has a similar scope as our section~\ref{sec:fit}.
The overall picture of our findings agrees with it, but there are some notable differences in the strategies. For example,
\begin{itemize}
 \item we use the $S_i, A_i$ basis instead of the observables suggested in \cite{Descotes-Genon:2013vna}. It is reassuring that both approaches lead to similar results.
 \item we include ATLAS and BaBar data for $B \to K^* \mu^+\mu^-$, which turns out to be important for the significance of the tension in $F_L$.
\item we do not observe a tension in $A_\text{FB}$ (or the alternative observable $P_2$) at low $q^2$. Also here, averaging all available experiments has a notable impact, as discussed in appendix~\ref{app:combination}.
 \item we include the $B\to K\mu^+\mu^-$ data. In spite of our inflated error bars at high $q^2$, the constraint turns out to be relevant and is crucial for limiting the allowed size of $|C_9^\text{NP}|$, explaining why we find that NP contributions in $C_9$ only cannot reduce the tensions in $S_5$ and $F_L$ completely.
\end{itemize}

In addition to the model-independent analysis, we also
studied the implications of the results for models of new physics. A naive dimensional estimate points towards a NP scale of several tens of TeV in the case of tree-level NP, or several TeV in the case of loop-level NP. 
In concrete well motivated models, like the MSSM or models with partial compositeness, as well as in models with flavour-changing $Z^\prime$ bosons, we find that new particles are typically required at the order of $\sim 1$~TeV:
\begin{itemize}
\item 
Models with flavour-changing $Z^\prime$ bosons can accommodate large NP in $C_9$ and $C_9'$. However, taking into account bounds from $B_s$ mixing, we find that the $Z^\prime$ bosons have to be light, of the order of few TeV at most. The strong bounds from searches for di-jet and di-lepton resonances at the LHC imply that for $Z^\prime$ masses around 1~TeV, the $Z^\prime$ couplings to first generation quarks have to be at least one order of magnitude smaller than the corresponding couplings of the SM $Z$ boson. 
\item
We showed that throughout the viable MSSM parameter space, contributions to $C_9$ are well below the values preferred by the model independent fit to the available data of $b\to s$ decays. 
The tensions in $B\to K^* \mu^+\mu^-$ can only be softened slightly in the MSSM, by NP contributing to the dipole operators $O_7^{(\prime)}$.
Such contributions can be generated in various ways: charged Higgs bosons with masses around 500~GeV; stops at or below 1~TeV; or gluinos and down-type squarks as heavy as 2~TeV, if generic squark flavour mixing is considered.   
\item
In models with partial compositeness, the tensions can also be softened by one-loop contributions to the dipole operators. Appreciable contributions to $C_9$ can be generated at tree level from heavy vector exchange but would require a significant degree of compositeness of one chirality of muons. The question whether such a scenario is viable deserves further study. In any case, it would also lead to sizable contributions to $C_{10}$ that would have to be cancelled by $Z$-mediated contributions.
\end{itemize}

Generically, most of the discussed NP explanations of the $B\to K^* \mu^+\mu^-$ anomalies can also lead to sizable CP asymmetries in $B\to K^* \mu^+\mu^-$, as shown in the examples of figure~\ref{fig:ACP}.
Improved results on the CP asymmetries $A_7$, $A_8$, and $A_9$
as well as on the CP averaged observables $F_L$, $A_\text{FB}$, $S_3$, $S_4$, $S_5$
and on other processes like $B\to K\mu^+\mu^-$
will be extremely useful to confirm the deviations currently observed, to establish the presence of new physics in $b \to s$ transitions, and to pin down its properties.

\section*{Acknowledgements}
We thank Christoph Bobeth for illuminating discussions and Nicola Serra for useful correspondence.
Fermilab is operated by Fermi Research Alliance, LLC under Contract No. De-AC02-07CH11359 with
the United States Department of Energy.
The research of D.S.\ is supported by the Advanced Grant EFT4LHC of the European Research Council (ERC), and the Cluster of Excellence {\em Precision Physics, Fundamental Interactions and Structure of Matter\/} (PRISMA -- EXC 1098). The research of W.A.\ was supported in part by Perimeter Institute for Theoretical Physics. Research at Perimeter Institute is supported by the Government of Canada through Industry Canada and by the Province of Ontario through the Ministry of Economic Development \& Innovation.

\appendix

\section{Data averages}\label{app:combination}

\begin{figure}[tp]
\centering
\includegraphics[width=0.46\textwidth]{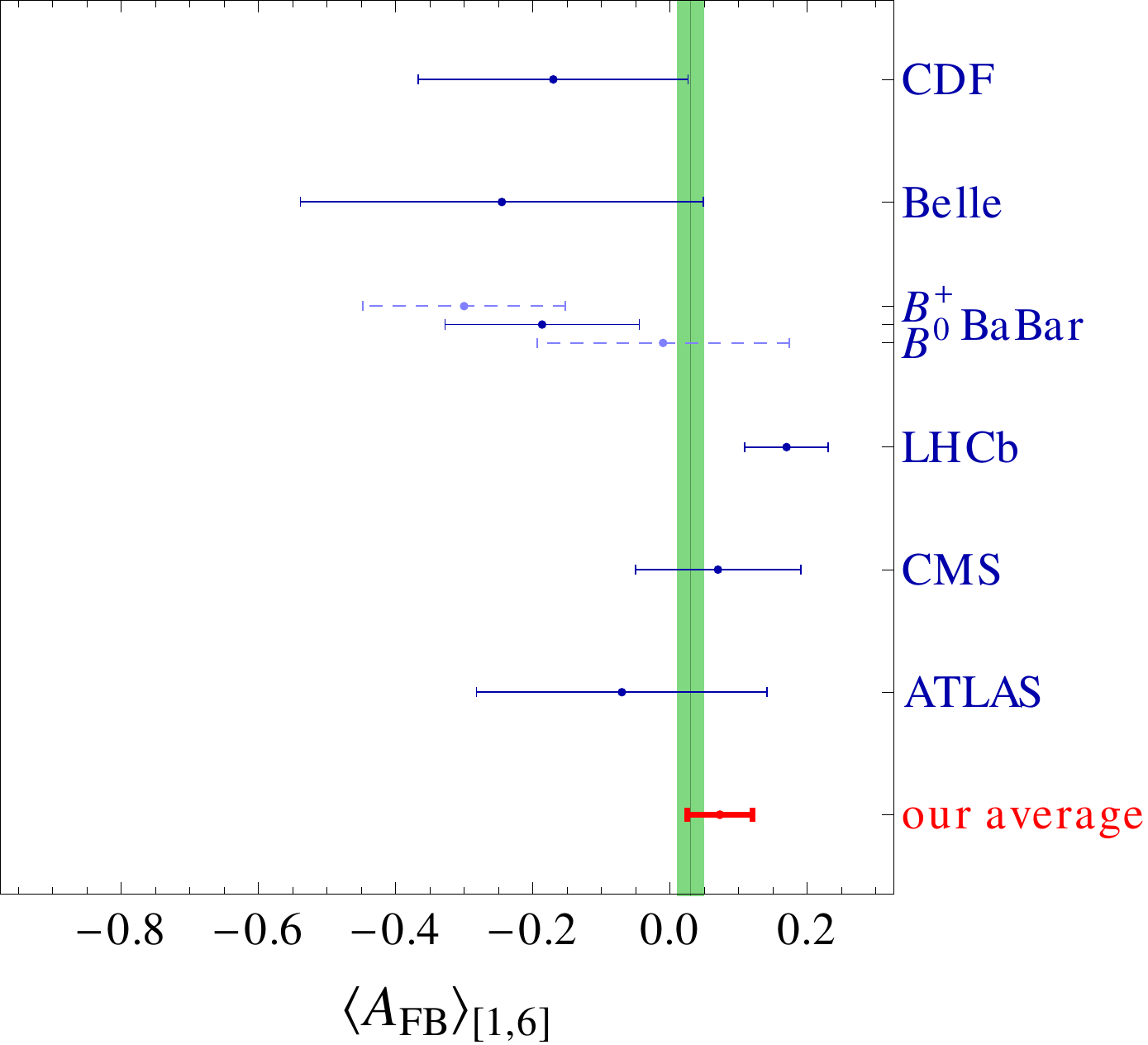} ~~~~~ \includegraphics[width=0.46\textwidth]{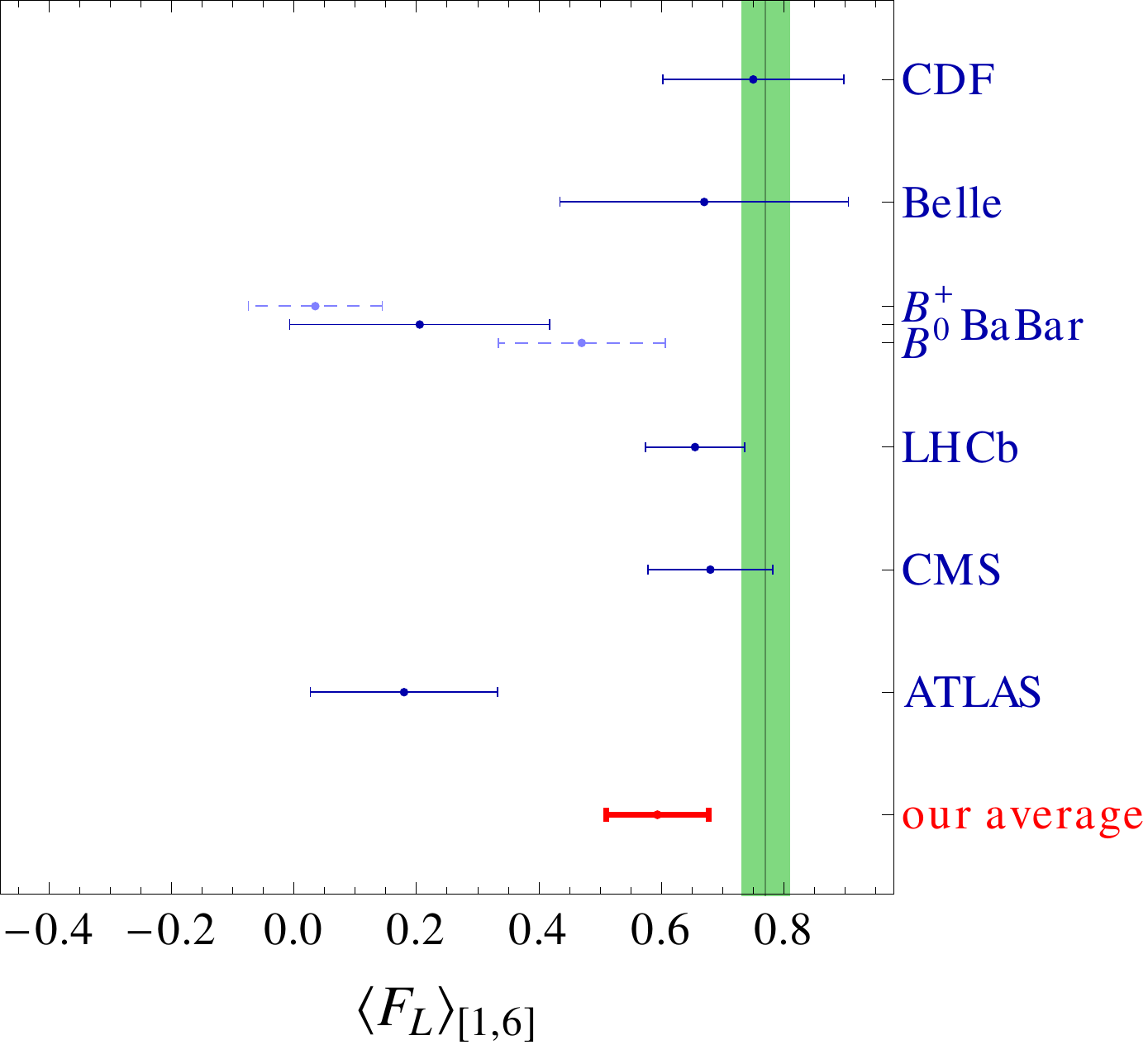}
\caption{Individual experimental results with $1\sigma$ uncertainties for $A_\text{FB}$ (left) and $F_L$ (right) in the $[1,6]$~GeV$^2$ bin, as well as our averages. The SM predictions with $1\sigma$ uncertainties are shown by the green bands.}
\label{fig:data}
\end{figure}

In this appendix we give more details on how we obtain the averages of experimental measurements of $B\to K^*\mu^+\mu^-$  observables~\cite{Wei:2009zv,Lees:2012tva,CDFupdate,Aaij:2013iag,Aaij:2013qta,ATLAS:2013ola,CMS:cwa} used in this analysis. 
As in~\cite{Altmannshofer:2012az}, we first symmetrize asymmetric statistical and/or systematic errors and then perform a weighted average of the symmetrized individual results.
While in many cases the obtained averages are dominated by the LHCb results, the averaging procedure leads to important shifts in the observables $F_L$ and $A_\text{FB}$ in the $[1,6]$~GeV$^2$ bin. This is illustrated in the plots of figure~\ref{fig:data}.
In the case of the BaBar $B\to K^*\mu^+\mu^-$ data at low $q^2$, the results for the charged and neutral modes show a significant difference. We therefore first average the charged and neutral mode results of BaBar, using the PDG averaging method, i.e. rescaling the uncertainty by a factor of $\sqrt{\chi^2}$. We then use this average and combine it with the available data from the other experiments.

In the case of $F_L$, we observe tensions between the data of the several experiments. In particular, in the $[1,6]$~GeV$^2$ bin, both BaBar and ATLAS data are significantly below the measurements of the other experiments and the SM prediction. Therefore, we rescale the uncertainty of our weighted average of $F_L$ by $\sqrt{\chi^2/N_\text{dof}}$. As shown in the right plot of figure~\ref{fig:data}, a tension with the SM prediction of $1.9\sigma$ remains. In the case of $A_\text{FB}$, the tension between the SM prediction and the LHCb data alone is softened considerably after data from the other experiments is taken into account (see left plot of figure~\ref{fig:data}).

A final comment is in order on the observables $S_4$ and $S_5$ that have only been measured by LHCb. Since ref.~\cite{Aaij:2013qta} does not provide data for these observables in the $[1,6]$~GeV$^2$ bin, we have reconstructed them using the data on $P_{4,5}'$ and $F_L$ as in table~\ref{tab:dict}, which is expected to be very close to a direct determination\footnote{N.~Serra, private communication.}.

In table \ref{tab:data}, we list the resulting experimental averages for the angular observables and confront them with our SM predictions.

\begin{table}
\renewcommand{\arraystretch}{1.2}
\centering
\begin{tabular}{ccccc}
\hline
Observable & $q^2$ & SM prediction & Experiment & Pull \\
\hline
 \multirow{3}{*}{$\langle F_L \rangle$} & ${[1,6]}$ & $0.77\pm 0.04$ & $0.59\pm 0.08$ & \boldmath $1.9$ \\
  & ${[14.18,16]}$ & $0.37\pm 0.19$ & $0.31\pm 0.06$ & $0.3$ \\
  & ${[16,19]}$ & $0.34\pm 0.24$ & $0.31\pm 0.05$ & $0.1$ \\
\hline
 \multirow{3}{*}{$\langle A_\text{FB}\rangle$} & ${[1,6]}$ & $0.03\pm 0.02$ & $0.07\pm 0.05$ & $0.9$ \\
  & ${[14.18,16]}$ & $-0.41\pm 0.12$ & $-0.47\pm 0.04$ & $0.4$ \\
  & ${[16,19]}$ & $-0.35\pm 0.13$ & $-0.36\pm 0.04$ & $0.1$ \\
\hline
 \multirow{3}{*}{$\langle S_3 \rangle$} & ${[1,6]}$ & $-0.00\pm 0.01$ & $0.03\pm 0.07$ & $0.4$ \\
  & ${[14.18,16]}$ & $-0.14\pm 0.08$ & $0.03\pm 0.09$ & $1.4$ \\
  & ${[16,19]}$ & $-0.22\pm 0.10$ & $-0.21\pm 0.09$ & $0.1$ \\
\hline
 \multirow{3}{*}{$\langle S_4 \rangle$} & ${[1,6]}$ & $0.10\pm 0.02$ & $0.14\pm 0.10$ & $0.4$ \\
  & ${[14.18,16]}$ & $0.29\pm 0.04$ & $-0.07\pm 0.11$ & \boldmath $2.8$ \\
  & ${[16,19]}$ & $0.31\pm 0.07$ & $0.16\pm 0.10$ & $1.1$ \\
\hline
 \multirow{3}{*}{$\langle S_5 \rangle$} & ${[1,6]}$ & $-0.14\pm 0.02$ & $0.10\pm 0.10$ &\boldmath $2.4$ \\
  & ${[14.18,16]}$ & $-0.35\pm 0.08$ & $-0.38\pm 0.13$ & $0.2$ \\
  & ${[16,19]}$ & $-0.26\pm 0.09$ & $-0.28\pm 0.09$ & $0.2$ \\
\hline
\end{tabular}
\caption{SM predictions confronted with experimental averages of $B\to K^*\mu^+\mu^-$ angular observables in the three $q^2$ bins. The pull is defined as $\sqrt{\Delta \chi^2}$. Details and references are given in the text.}
\label{tab:data}
\end{table}

\section{Loop functions}\label{loopfunctions}

In this appendix we collect the loop functions that appear in the discussion of the NP contributions to the Wilson coefficients $C_9^{(\prime)}$ and $C_7^{(\prime)}$ in sections~\ref{sec:MSSM} and~\ref{sec:PC}.

The loop functions entering MSSM contributions to the vector coefficients $C_9$ and $C_9^\prime$ read
\begin{equation}
f_9^{H^\pm}(x) = -\frac{2(38-79x+47x^2)}{9(1-x)^3} - \frac{4(4-6x+3x^3)\log x}{3(1-x)^4} ~~~\xrightarrow{x \to 1} 1 ~,
\end{equation}
\begin{equation}
f_9^{\tilde H^\pm}(x) = -\frac{2(52-101x+43x^2)}{21(1-x)^3} - \frac{4(6-9x+2x^3)\log x}{7(1-x)^4} ~~~\xrightarrow{x \to 1} 1 ~,
\end{equation}
\begin{equation}
f_9^{\tilde g}(x) = \frac{5(1-5x+13x^2+3x^3)}{3(1-x)^4} + \frac{20x^3 \log x}{(1-x)^5} ~~~\xrightarrow{x \to 1} 1 ~,
\end{equation}
\begin{equation}
f_9^{\tilde W}(x) = -\frac{10(22-38x+7x^2+3x^3)}{3(1-x)^4} - \frac{10(3-9x^2+4x^3) \log x}{(1-x)^5} ~~~\xrightarrow{x \to 1} 1 ~,
\end{equation}
\begin{equation}
f_9^\text{box}(x,y) = \frac{12(x-2y+xy)}{(1-x)(y-x)(1-y)^2} - \frac{12x^2 \log x}{(1-x)^2(x-y)^2} + \frac{12 y(2x-y+y^2)\log y}{(x-y)^2(1-y)^3} ~~~\xrightarrow{x,y \to 1} 1 ~.
\end{equation}

The loop functions that are relevant for the MSSM contributions to the dipole coefficients $C_7$ and $C_7^\prime$ read
\begin{equation}
 f_7^{H^\pm}(x) = \frac{3(5x-3)}{7(1-x)^2} + \frac{6(3x-2)}{7(1-x)^3} \log x ~~~\xrightarrow{x \to 1} 1 ~,
\end{equation}
\begin{equation}
 f_7^{\tilde H^\pm}(x) = \frac{6(7x-13)}{5(1-x)^3} + \frac{12(2x^2-2x-3)}{5(1-x)^4} \log x ~~~\xrightarrow{x \to 1} 1 ~,
\end{equation}
\begin{equation}
 f_7^{\tilde g}(x) = \frac{10(1+10x+x^2)}{(1-x)^4} + \frac{60x(1+x)}{(1-x)^5} \log x ~~~\xrightarrow{x \to 1} 1 ~.
\end{equation}

The function entering the Higgs loop contribution to $C_7$ and $C_7^\prime$ in models with partial compositeness reads
\begin{equation}
 f_7^h(x) = \frac{x(x^2-4 x+2 \log (x)+3)}{(x-1)^3} ~~~\xrightarrow{x\to\infty}1  ~.
\end{equation}

\bibliographystyle{JHEP}
\bibliography{bsll}

\end{document}